\documentclass[hyper]{JHEP3}

\input{epsf}
\usepackage{epsfig}
\usepackage{amssymb}
\usepackage{amsfonts}
\usepackage{amsbsy}
\usepackage[all,2cell]{xy}

\usepackage{amsmath}

\usepackage{amssymb,amscd}
\usepackage{mathrsfs}
\usepackage{amsmath,amsthm}

\def\be{\begin{equation}}
\def\ee{\end{equation}}

\def\Hom{{\rm Hom}}

\def\Mat{{\rm Mat}}
\def\mod{{\rm mod}}
\def\Tr{{\rm Tr}}

\def\IC{\mathbb{C}}

\def\IQ{\mathbb{Q}}
\def\IR{{\mathbb{R}}}

\def\IZ{{\mathbb{Z}}}

\def\CB{{\cal B}}
\def\CC {{\cal C}}

\def\CN {{\cal N}}

\def\CD {{\cal D}}

\def\CE {{\cal E}}

\def\CH {{\cal H}}

\def\CB {{\cal B}}

\def\CT {{\cal T}}

\def\half{\frac{1}{2}}

\def\one{{\hbox{ 1\kern-.8mm l}}}

\def\p{\partial}

\def\be{\bar{e}}

\def\half{\frac{1}{2}}

\def\half{\frac{1}{2}}

\def\ZZ{\mathbb{Z}}

\def\be{ \begin{equation} }
\def\ee{ \end{equation}}

\def\fg{ \mathfrak{g}}

\def\fa{\mathfrak{a}}

\def\fd{\mathfrak{d}}
\def\fe{\mathfrak{e}}

\def\fg{\mathfrak{g}}

\def\fl{\mathfrak{l}}

\def\fs{\mathfrak{s}}
\def\ft{\mathfrak{t}}

\def\fs{\mathfrak{s}}
\def\ft{\mathfrak{t}}
\def\fu{\mathfrak{u}}

\def\fB{\mathfrak{B}}

\def\fI{\mathfrak{I}}
\def\fL{\mathfrak{L}}

\def\fN{\mathfrak{N}}

\def\I{{\rm i}}

\def\Aut{{\rm Aut}}

\title{An Uplifting Discussion of T-Duality}

\author{Jeffrey A. Harvey$^1$ and Gregory W.~Moore$^2$ \\
$^1$ Enrico Fermi Institute and Department of Physics, University of Chicago\\
$~~ $5640 Ellis Ave., Chicago IL 60637, USA \\
$^2$ NHETC and
$~~$Department of Physics and Astronomy, Rutgers University \\
$~~$126 Frelinghuysen Rd., Piscataway NJ 08855, USA\\
\\
{\tt j-harvey@uchicago.edu, gmoore@physics.rutgers.edu } }

\abstract{It is well known that string theory has a T-duality symmetry relating circle compactifications of large and small radius.
This symmetry plays a foundational role in string theory.
We note here that while T-duality is order two acting on the moduli space of compactifications, it is order four in its action on the conformal field theory state space. More generally,
involutions in the Weyl group $W(G)$ which act at points of enhanced $G$ symmetry have canonical lifts to order four elements of $G$, a phenomenon first investigated by J. Tits in the mathematical literature on Lie groups and generalized here to conformal field theory. This simple fact has a number of interesting consequences. One consequence is a reevaluation of a mod two condition appearing in asymmetric orbifold constructions. We also
briefly discuss the implications for the idea that T-duality and its generalizations should be thought of as discrete gauge symmetries in spacetime.
\newline
\newline
 \today }

\begin{document}

\section{Introduction}\label{sec:Introduction}

This paper discusses the structure of,  and consistency conditions for, group actions on two-dimensional conformal field theories (CFTs)
defined by sigma models with toroidal target spaces. Such models are important building blocks in string theory. For simplicity
we will discuss the bosonic string, but our considerations should have generalizations to heterotic and type II superstrings.  We will see that some standard results in the literature have  minor inaccuracies and we will
indicate how these can be corrected. Some of the implications for general statements about string theory are also discussed. The central
point of this paper is easily stated: Some toroidal compactifications have symmetries of the lattice of momenta and winding
beyond the trivial symmetry of reflection in the origin. Sometimes, these lattice symmetries act projectively on the CFT state space. When discussing symmetries of CFTs or constructing orbifolds this subtlety can be of some importance. This should come as no surprise: It is extremely common for symmetries of physical systems to be realized projectively on quantum Hilbert spaces. The surprise, perhaps, is the extent
to which this elementary point has been overlooked in the literature.

There are two ways to detect the need for a projective action. One is based on  modular covariance and is explained in
section \ref{subsec:TechSumm} below. The second is based on non-abelian symmetry and is more easily explained.
The moduli space of toroidal compactifications is well known to have points of enhanced symmetry.  For example there are
points where the moduli space has an action of the Weyl group $W(G)$
where $G$ is a Lie group whose Lie algebra is simply laced, that is of type $A_n$, $D_n$, $E_6$, $E_7$ or $E_8$ (ADE).
It is often assumed that  the group that acts  on the moduli space at points of enhanced symmetry is also the group that acts on the CFT space.
However this is often not  the case.  As we will explain later, for some $G$ the Weyl group $W(G)$ does not lift to a subgroup of $G$  whose
action on $T$ by conjugation is isomorphic to that of $W(G)$, and which is isomorphic to $W(G)$. Rather one must choose a lift to a group $\widetilde W(G)$ where the order $2$ elements that generate $W(G)$ lift to elements of order
$4$ in order to produce the desired action by conjugation. This subtlety is relevant for
the theory of orbifolds since in an orbifold one gauges a subgroup of the group of automorphisms
of the CFT, not a group of automorphisms of Narain moduli space.  The orbifold construction plays
an  important role in string theory and Conformal Field Theory (CFT)  \cite{Dixon:1985jw,Dixon:1986jc,Polchinski:1998rq} and we will see
that this subtlety has implications for string-theoretic model building.
For example we will show that a larger class of asymmetric orbifold
constructions is allowed than is sometimes thought to be the case.
The reason for this is that in the past some models have been discarded
because they do not satisfy a certain mod-two consistency condition stated
in the second part of equation $(2.6)$ of   \cite{Narain:1986qm}.
What is less well-known is that this mod two condition was retracted in \cite{NSVII},
just above equation $(3.3)$ , where it
was suggested that one should double the order of the the group element.
That reinterpretation is closely related to the discussion we give in this paper.


The outline of this paper is as follows. In section \ref{subsec:TechSumm} we review the construction of toroidal
CFTs and summarize the main points of the paper in more technical language than used here. The rest of the paper
is a more leisurely exposition of this technical summary. In particular, we begin the story in section three
by considering the basic example of the $c=1$ Gaussian model at the self-dual
point with affine level one $SU(2)_L \times SU(2)_R$ symmetry. We explain that the order two T-duality transformation which fixes this
point in the moduli space lifts to an order four element of $SU(2)$, both from the point of view of group theory and from the point of view
of modular covariance. We then use this point of view to analyze the consistency of orbifolds of products of self-dual Gaussian models by this order four lift of T-duality. In section four we extend our analysis of lifting of Weyl group symmetries to points with enhanced ADE symmetry and provide some illustrative examples. In section five we return to the $SU(2)$ analysis and explain how
to understand the order four action of $T$-duality by carefully evaluating how the symmetry acts on vertex operators.
    Section six is devoted to a general analysis of consistency
conditions of asymmetric orbifolds by the lifts of involutions of the Narain lattice and in section 7 we make some comments about more general
asymmetric orbifolds.  In CFT and Vertex Operator Algebras one must deal with an abelian extension of the Narain lattice and here we are also interested in the associated extension relating the automorphism group of the lattice to the automorphism group of its extension. The first appendix discusses the required mathematics. The remaining three appendices contain material on the transformation of orbifold boundary conditions under modular transformations, our conventions for theta functions, and a brief summary of the mathematical structure of lifts of the Weyl groups of compact simple Lie groups.

\bigskip
\bigskip
\noindent
\textbf{\emph{Note added for v3}}: In the first two versions of this paper posted on the arXiv we
claimed that it is strictly necessary to modify the standard $\IZ_2$-valued cocycles in
vertex operator algebras to $\IZ_4$-valued cocycles in order to understand the nontrivial lifting
of $T$-duality discussed throughout the paper. This claim is erroneous. It was pointed out to us by the referee
that one can perfectly well use the standard cocycles with a suitable modification of the lifting function
See  equation \eqref{eq:LiftingFunction} and note added below for further details. 
  We thank the referee for insisting on this point.

\section{Technical Summary Of Results}\label{subsec:TechSumm}

\subsection{Review Of Toroidal CFT And T-Duality}

In order to state our results with more precision we first recall the
essential elements of toroidal conformal field theories.
As is well-known, two-dimensional CFTs of free  scalar fields with toroidal target space
are not isolated. There are actually two constructions of these CFTs which we may call the
\emph{vertex operator algebra (VOA) construction} and the \emph{sigma model construction}. Each construction
has its advantages, and each construction leads to a parameter space of conformal field theories which
requires taking a quotient to obtain the moduli space of toroidal conformal field theories.

In the vertex operator algebra construction we begin with an embedding of the unique even unimodular lattice
$II^{d_L,d_R}$, of signature $(+^{d_L}, -^{d_R})$  into a fixed real quadratic space $V$ equipped with
projection operators to a positive definite space of dimension $d_L$ and a negative definite space of
dimension $d_R$. We can identify $V$ with the standard space  $\IR^{d_L;d_R}$ with diagonal quadratic
form and projections onto the first $d_L$ and last $d_R$ coordinates, respectively.
 (The semi-colon is meant to remind us that this space comes with definite projection operators.)
 We assume that  $d_L, d_R>0$ and $d_L-d_R = 0~\mod ~ 8$.
We denote the image of $II^{d_L,d_R}$ by $\Gamma \subset \IR^{d_L;d_R}$. The moduli space of
such embeddings is the homogeneous space
\be\label{eq:EmbedDef}
\fL := \CT \backslash O(V)
\ee
where $\CT \cong \Aut(II^{d_L,d_R})$ is the T-duality group, usually written as $O(d_L,d_R;\IZ)$. (The latter notation
is less precise, as it presupposes an integral quadratic form, but it is standard, so we will use it.
 With the same understanding it is also common to write $O(V)$ as $O(d_L,d_R; \IR)$.)
Now for each $\Gamma \in \fL$ we can construct a 2d CFT $\CC_{\Gamma}$ as follows.
The vector space of left-moving creation oscillators
(for any fixed positive integer frequency) can be identified with $V_L \otimes \IC$, where $V:=\Gamma\otimes \IR\cong \IR^{d_L;d_R}$
and $V_L$ is its left-moving projection. Similarly the vector space of the right-moving creation
operators (for a fixed frequency) is $V_R \otimes \IC$. In these terms the CFT state space can be written as:
\be\label{eq:CFT-statespace}
\CH_\Gamma = S^\bullet(\oplus_{n>0} q^n V_L\otimes \IC) \otimes S^\bullet(\oplus_{n>0} \bar q^n V_R\otimes \IC)
\otimes \IC[\Gamma] \, .
\ee
Here $S^\bullet(\oplus_{n>0} q^n V_L\otimes \IC)$ denotes the symmetric algebra of the left-moving
creation oscillators with positive frequency. The factor $q^n$ is meant to indicate the space with frequency $n$.
Similarly,   $S^\bullet(\oplus_{n>0} \bar q^n V_R\otimes \IC)$ is
the symmetric algebra of the right-moving creation oscillators. $\IC[\Gamma]$ is the group algebra of the
Narain lattice. As a vector space it is a direct sum of lines $L_p \cong \IC$, one line associated to each momentum vector $p\in \Gamma$.
The space $\CH_{\Gamma}$ can be given the structure of a (in general, nonholomorphic) vertex operator algebra,
although the details require some care, as recalled in section \ref{subsec:CocycleReview}.  Note that we therefore
have a bundle of CFTs over $\fL$, with fiber $\CH_{\Gamma}$ above $\Gamma \in \fL$.

Different embeddings
$\Gamma, \Gamma' \subset V$ can lead to isomorphic conformal field theories. This happens if they are related
by the action of the subgroup $O(d_L) \times O(d_R)$ of $O(d_L,d_R;\IR)$. For example, the Hamiltonian
$H = \half p_L^2 + \half p_R^2 + H^{\rm osc}$ commutes with this group. Therefore the true moduli space of
conformal field theories is the quotient, known as Narain moduli space, and  can be identified with the
double coset:
\footnote{We are actually being somewhat sloppy here from a mathematical viewpoint.
(Most physicists will want to skip this footnote.)
The ``Narain moduli space'' is an orbifold, and is more properly regarded as a global stack where the
automorphism group of objects is always a finite group. However,
it is not really the moduli stack of toroidal conformal field theories. In the latter
stack, the automorphism group of an object will include continuous groups at, for example,
the points of enhanced A-D-E symmetry, while in the Narain moduli stack the automorphism group
of the A-D-E points is a finite group $F(\Gamma(\fg))$ discussed at length below. The
moduli stack of conformal field theories maps to the Narain moduli space.}
\be\label{eq:NarainSpace}
\CN :=   \CT   \backslash O(d_L,d_R;\IR)/ O(d_L) \times O(d_R) \, .
\ee

We now recall briefly the sigma model construction. Since we do not wish to
enter into the subtleties of quantizing the self-dual field
 we will   limit considerations to theories with  $d=d_L=d_R$. In this case
 one may easily write an action for the sigma model using the data of a
 flat metric, $G$, and $B$-field on the torus. Thus, the moduli space of
 sigma model data is
\be\label{eq:SigModel-Moduli}
\fB := \{ E= G +B \vert G= G^{tr}>0 \quad \& \quad B= - B^{tr} \} \subset \Mat_{d\times d}(\IR)   \, .
\ee
This space is isomorphic to $O(d,d;\IR)/O(d) \times O(d)$ as a smooth manifold.
To illustrate we use a construction going back to \cite{Narain:1986am,Ginsparg:1986bx}
(but here slightly modified from the original).
Choose   two invertible $d\times d$ matrices $e_1, e_2$ so that
$e_1 e_1^{tr} = e_2 e_2 ^{tr}= G^{-1} $. Note that $e_1$ and $e_2$ are defined up to
right action by an $O(d)$ matrix. Now define the $2d \times 2d $ matrix:
\be
\CE : = \begin{pmatrix}  \half e_1 & \half e_2   \\
 E^{tr} e_1  & -  E e_2  \\
  \end{pmatrix}
\ee
The reader can readily check that this solves
\be\label{eq:InnProds}
\CE Q_0 \CE^{tr} = Q
\ee
where
\be
Q_0 = \begin{pmatrix} 1_{d} & 0 \\ 0 & -1_d \\ \end{pmatrix}
\qquad\qquad
Q = \begin{pmatrix}
0 & 1_d \\
1_d & 0 \\ \end{pmatrix}
\ee
Since $Q$ and $Q_0$ are similar the space of matrices solving \eqref{eq:InnProds}
is smoothly isomorphic to $O(V)$. Modding out by the right action on $\CE$ of
$O(d_L) \times O(d_R)$ produces, on the one hand, the space $\fB$ and on the
other hand, the coset $O(d_L,d_R;\IR)/\left( O(d_L) \times O(d_R) \right)$.

Now, quite similarly to the case of the bundle of CFT state spaces over $\fL$ we
can likewise produce a bundle of state spaces $\CH$ over $\fB$.
\footnote{Once again we are being somewhat sloppy from a strictly mathematical
point of view. Canonical quantization only provides a projective Hilbert space
because it is based on a choice of vacuum \underline{line}, rather than a choice of vacuum
\underline{state}.
Thus, what is canonically defined is a bundle of projective Hilbert spaces. One might
dismiss this subtlety because $\fB$ is a contractible space. However, the space is not
equivariantly contractible, so this issue will, doubtless, be of some importance in
sorting out the issues of T-duality as a gauge symmetry mentioned below.}
We denote the fiber over $E$ by $\CH_E$. It is produced by canonical quantization,
and in the process of quantization one finds - after fixing the gauge for $O(d_L)\times O(d_R)$ -
 that the lattice of zero frequency
momentum and winding modes is the embedded lattice in $\IR^{d;d}$ generated by integer
combinations of the rows of $\CE$. By equation \eqref{eq:InnProds} this is an even
unimodular lattice and hence defines an element of $\fL$. Tracing back the change of
basis to an action of $\CT$ on $\fB$ produces the familiar left action of $\CT$ on $\fB$
via  fractional linear transformations  of $E$.  The whole construction can be summarized in
the diagram:
\be\label{eq:DiamondDiag}
\xymatrix{
 & O(V)\ar[dl] \ar[dr] &   \\
 \fL\ar[dr] &  &  \fB  \ar[dl] \\
  &  \CN &  \\
}
\ee
where the left-hand path is the vertex operator algebra construction and the right-hand path is
the sigma-model construction.

\subsection{The Enhanced Symmetry Locus}

The space $\CN$ has an important subspace $\CN^{\rm ESP}$ of points with \emph{enhanced symmetry} that will be important in this paper.
To define the enhanced symmetry locus first note that every
embedded lattice $II^{d_L,d_R} \hookrightarrow \Gamma \subset \IR^{d_L;d_R}$ has an automorphism
corresponding to $p \to - p$. We will call this the \emph{trivial involution}.
Note that this is always in $O(d_L) \times O(d_R)$ for any embedding. However, on a positive
codimension subvariety $\fL^{\rm ESP} \subset \fL$ the there will be nontrivial automorphisms of the CFT.
To be precise, define the group:
\be\label{eq:CrystalGroup}
F(\Gamma):= \Aut(\Gamma) \cap \left( O(d_L) \times O(d_R) \right) \, .
\ee
Note that this group is both discrete and compact and hence is a finite group. The locus
$\fL^{\rm ESP} \subset \fL$ is defined to be the set of embeddings such that $F(\Gamma)$ is
 strictly larger than the central  $\IZ_2$ subgroup generated by the trivial involution.
In a neighborhood of $\fL^{\rm ESP}$ the action of $O(d_L) \times O(d_R)$ has fixed points,
producing a complicated subvariety $\CN^{\rm ESP}$ of orbifold singularities where the
orbifold group is, generically,  $F(\Gamma)/\IZ_2$.
Thanks to equation \eqref{eq:DiamondDiag} we know there is a corresponding locus $\fB^{\rm ESP}\subset \fB$
where a finite subgroup of $\CT$ acts with fixed points.
\footnote{In the interest of technical accuracy we note  that \eqref{eq:CrystalGroup} for different
$\Gamma$ projecting to the same point in $\CN$ will be conjugate groups. Similarly, we will often loosely
speak of $F(\Gamma)$ when working with a point $E\in \fB$. What is meant here is that one fixes the
$O(d_L)\times O(d_R)$ gauge by choosing inverse vielbeins $e_1, e_2$ as above and then constructs a particular
$\Gamma$ using the integer span of the rows of $\CE$. }

The orbifold singularities at points $[\Gamma] \in \CN$ signal the presence of nontrivial automorphisms
of the conformal field theory  $\CC_\Gamma$ parametrized by $\Gamma$.
 In the string theory literature it is commonly assumed
that $F(\Gamma)$ can be identified with a group of automorphisms of the CFT $\CC_\Gamma$,
but - \emph{and this is the central point of this paper} -  this is not always the case, and the distinction
between $F(\Gamma)$ and $\Aut(\CC_{\Gamma})$  can be important.
How can this happen? To explain this point we note that   the group $F(\Gamma)$ acts on the Narain lattice $\Gamma$ and that action
extends linearly to the vector space $V = \Gamma \otimes \IR$ and commutes with the left-moving and
right-moving projectors. Therefore it acts naturally on the left-moving and right-moving oscillators.
However, we must \underline{also} determine the action on $\IC[\Gamma]$ and here is where a subtlety can arise. In physical terms, we choose a generating vector for each $L_p$ (it is the ground state in
the momentum sector $p$)  and denote it by   $\vert p \rangle$.  In the literature one commonly finds the claim that we can choose a basis of momentum states $\vert p\rangle$ so that,
for all $g\in F(\Gamma)$, there is an operator $U(g)$ on $\CC_{\Gamma}$
such that
\footnote{In general it is also possible to include the action by
\emph{shift vectors}. Group elements are labeled by $(g,s)$
where $s \in \Gamma \otimes \IQ $ is known as a shift vector and we modify the
action \eqref{eq:NaiveAct} by the (equally naive) action:
\be\label{eq:NaiveActShift}
U(g,s) \vert p \rangle = e^{2\pi \I p \cdot s} \vert g \cdot p \rangle.
\ee
We are not trying to be comprehensive and will, for the most part, ignore
the inclusion of shift vectors in this paper. However, the incorporation of
shift vectors will play a role in some examples below.}
\be\label{eq:NaiveAct}
U(g) \vert p \rangle =   \vert g \cdot p \rangle.
\ee
While this is commonly assumed, it turns out that it is, in general, not
consistent with the non-abelian global symmetry of special CFTs associated with
special points in $\CN$. It is also inconsistent with the same non-abelian global symmetry of the Operator Product Expansion (OPE),
given  the state-operator correspondence. More generally, at points where $F(\Gamma)$ is nontrivial
it is, in general, inconsistent with \emph{modular covariance}. (This term is explained below.)

\subsection{Non-Abelian Symmetry}

The simplest example of a conflict between  equation \eqref{eq:NaiveAct} and
non-abelian global symmetry is the
Gaussian model at the self-dual radius.    This CFT is, famously, equivalent to
 the level 1 $SU(2)$ WZW model \cite{Frenkel:1980rn,Segal:1981ap,Witten:1983ar}.
\footnote{The following transparent example arose in discussions with N. Seiberg and has been
quite important to our thinking. After submitting v1 of this paper it was pointed out to us that
 this particular example has been previously discussed by
 Aoki, D'Hoker, and Phong \cite{Aoki:2004sm}. Related works include \cite{Satoh:2015nlc,Satoh:2016izo}.}

In order to avoid confusion it is important to specify precisely what the symmetries are of the Gaussian model and why we focus on a particular element that we call T-duality.
At the self-dual point the Gaussian model
has $su(2)_L \times su(2)_R$ affine symmetry. Focus for the moment on the $su(2)_R$ symmetry. There are two order $2$ automorphisms of the $su(2)$ Lie algebra
which are commonly used in the string theory literature. The first, a $\ZZ_2$ twist, acts on the currents as $\tilde J^3 \rightarrow - \tilde J_3$,
$\tilde J^\pm \rightarrow \tilde J^\mp$. In the Frenkel-Kac-Segal construction of affine $su(2)_R$ this is implemented by the transformation $X_R \rightarrow - X_R$. Denote this transformation by $\sigma_R$. Clearly we can do the same thing but on holomorphic (left-moving) degrees of freedom. Denote
this transformation by $\sigma_L$.  In addition to these ``twist" transformations we can consider ``shifts." An order $2$ shift on the anti-holomorphic degrees of freedom at the self-dual radius acts on the bosonic coordinate as $\tilde X_R \rightarrow  \tilde X_R + \pi / \sqrt{2}$ and takes
$\tilde J_3 \rightarrow \tilde J_3$, $\tilde J^\pm \rightarrow - \tilde J^\pm$. There is an analogous order two symmetry acting on holomorphic
degrees of freedom. Let us denote these by $S_R, S_L$.
In the notation of the previous paragraph we take T-duality to be $\sigma_R$. This is a symmetry which exists only at the self-dual radius.
The usual symmetric $\ZZ_2$ action used to construct the $\ZZ_2$ symmetric orbifold of the Gaussian model is $\sigma_L \sigma_R$ and exists
at any radius.  An alternative version of $T$-duality at the self-dual radius proposed in \cite{Hellerman:2006tx}  is $ S_L \sigma_R$.
However it is easy to check that $S_L = g^{-1} \sigma_L g $ for $g \in SU(2)_L$.  Now $SU(2)_L$ is a global symmetry of the CFT at the self-dual
radius. Hence any two operators which are conjugate in $SU(2)_L$ will lead to identical physical predictions. In particular, any computation
involving the $\ZZ_2$ operator $ S_L \sigma_R$ will give physically identical results to a computation using the $SU(2)_L$ conjugate
operator $ \sigma_L \sigma_R$ which is the symmetric $\ZZ_2$ operator. Since this alternate ``T-duality" is just the symmetric $\ZZ_2$
symmetry in disguise it is not surprising that the $\ZZ_2$ orbifold  by it is consistent and that transformations acting on the CFT state space are order $2$. However
while the orbifold by this symmetry is consistent, it simply reproduces the usual symmetric orbifold of the Gaussian model. From now on we use T-duality to refer
to the left-right asymmetric symmetry $\sigma_R$ and its generalizations.

When acting on the  currents of the model, T-duality acts trivially on (say) the left-moving currents
but acts as a   $180$ degree rotation on the right-moving currents. On the other hand there are states
in the model that transform as the tensor product of left- and right-moving
 spinor representations of $SU(2)_L \times SU(2)_R$. Therefore, in order to define an action on
 the Hilbert space we must lift the $180$-degree rotation in the right-moving $SO(3)$ to an element in the
 right-moving $SU(2)$. This lift to $SU(2)$ is clearly of order four. This phenomena generalizes to the standard enhanced symmetry loci in
$\CN$ associated with semi-simple simply-laced Lie algebras. If $\fg$ is of full rank (and $d_L = d_R$) these are isolated points
defined by
\be\label{eq:ESP-DEF}
\Gamma(\fg) := \{ (p_L;p_R) \in \Lambda_{wt}(\fg) \times \Lambda_{wt}(\fg) \vert  p_L - p_R \in \Lambda_{rt}(\fg) \}.
\ee
The corresponding CFT,  $\CC(\fg):=\CC_{\Gamma(\fg)}$ has $\widetilde{LG}^{(1)}_L \times \widetilde{LG}^{(1)}_R$ (dynamical) symmetry, where
$G$ is the compact simply connected Lie group with Lie algebra $\fg$ and $\widetilde{LG}^{(1)}$ is the level one $U(1)$
central extension of the loop group $LG$.  In particular
it has an action of  $G_L \times G_R$ corresponding to the constant loops. On the other hand, the crystallographic
group $F(\Gamma(\fg))$ certainly contains
\be\label{eq:ESP-Crystal}
 W(\fg)_L \times W(\fg)_R
\ee
as a subgroup, where $W(\fg)$ is the Weyl group of $\fg$.

We must stress that $W(\fg)$ is \underline{not} a subgroup of $G$.
This seemingly fastidious point will actually turn out to be important. This point has been noted before in
the physics literature, see \cite{Schellekens:1987ij} where some of the material below is also discussed.
Quite generally, the Weyl group $W(\fg)$ is defined as follows. Choose a maximal torus $T\subset G$
and define the normalizer group $N(T):= \{ g \in G \vert  g T g^{-1} = T \}$. Of course $T \subset N(T)$,
and in fact the conjugation action by $T$ fixes every element pointwise, since $T$ is abelian. However,
the definition of $N(T)$ only requires conjugation to fix $T$ setwise, and
there are other elements of $G$ which conjugate the maximal torus to itself but do not fix every element of
$T$. In fact $T$ is a normal subgroup of $N(T)$ and the Weyl group is \underline{defined} as the quotient
\footnote{In fact, $W(\fg)$ is intrinsically associated to the Lie algebra $\fg$ and
 does not depend on which Lie group $G$ with Lie algebra $\fg$ we choose. Indeed, there are
 other, equivalent,   definitions of the Weyl group which only make direct use of the root system of $\fg$,
 rendering this property obvious. We have chosen to emphasize the relation to the Lie group since it fits
 best with the main point of the paper. }
\be\label{eq:WG-Def}
W(\fg) := N(T)/T \, .
\ee
Thus $W(\fg)$ is not a subgroup of $G$ but rather, it is a quotient of a subgroup of $G$ - that is, it is a
subquotient of $G$.
It follows from \eqref{eq:WG-Def} that $W(\fg)$ fits in an exact sequence
\be\label{eq:WG-Seq}
1 \rightarrow T \rightarrow N(T) \rightarrow W(\fg) \rightarrow 1 \, .
\ee
One can show that there are (many) discrete subgroups $\widetilde W \subset N(T) \subset G$
together with a homomorphism $\pi: \widetilde W \to W$ such that the conjugation action of $g\in \widetilde W $ on
the Cartan subalgebra $\ft \subset \fg$ is identical to the Weyl group action of $\pi(g)$. Such a subgroup
$\widetilde W \subset G$ is called a \emph{lift of $W$}. In some cases ($G=SU(2)$ is a case in point) there
is no lift isomorphic to $W$. Thus, at enhanced symmetry points of the form $\CC(\fg)$, many discrete automorphism groups
$\widetilde W_L \times \widetilde W_R$  of
$\CC(\fg)$ induce the action of $F(\Gamma(\fg))$ on oscillators and momenta. Moreover, in some
cases, no such lifting group is isomorphic to $W_L \times W_R$.

Since states are in one-one correspondence with operators in a CFT we expect that there will
be an analogous story for the automorphisms of the vertex operator algebra, and indeed this is the
case. It is well-known that the naive expression $V^{\rm naive}(p) = : e^{\I p \cdot X}:$ for the vertex operators
associated to momentum vectors must be modified by ``cocycle factors.'' In section  \ref{sec:CocycleComments}
we explain how this works for  $\fg = \fs\fu(2)$.
%
%

\subsection{Modular Covariance}

Now let us turn to conflicts between \eqref{eq:NaiveAct} and modular covariance.
We first explain the term ``modular covariance.''  Quite generally,
if $J$ is a global symmetry of a CFT $\CC$ then we can ``couple $\CC$ to external $J$ gauge fields.''
What this means is that, if the worldsheet is $\Sigma$ then we consider a principal $J$-bundle
over $\Sigma$ endowed with connection and couple the connection to the global symmetry currents of $\CC$.
\footnote{This must be distinguished from gauging the $J$ symmetry (as one does to form an orbifold).
In that case one sums over isomorphism classes of $J$-bundles with connection.}
If $J$ is a discrete group then there is a unique connection on the principal bundle and coupling
to the currents means imposing twisted boundary conditions by elements $g_a\in J$ around a set of generating
cycles of $\pi_1(\Sigma, *)$. The diffeomorphism group acts on this picture relating
different twisted partition functions. In the case of a torus we choose two commuting elements $g_s, g_t$
for twisting around a choice of $A$ and $B$ cycles and form the partition function $Z(g_t, g_s; \tau)$.
The ``modular covariance'' constraint is the statement that (see Appendix \ref{bctrans}):
\footnote{The term ``modular covariance'' was used in a slightly
different way in \cite{Aoki:2004sm} where the term is used for the
same identity but with the phase $e^{i \phi(\gamma)}$ put to one.   }
\be\label{eq:Mod-Cov}
Z(g_t, g_s; \frac{a \tau + b }{c\tau + d} ) = e^{\I \phi(\gamma) } Z(g_s^{-b} g_t^{d}, g_s^{a} g_t^{-c}; \tau)
\qquad \qquad \forall \gamma = \begin{pmatrix} a & b \\  c & d \\  \end{pmatrix} \in SL(2,\IZ)
\ee
where $e^{\I \phi(\gamma)}$ is some $U(1)$-valued function of $\gamma$, reflecting the possibility
of a modular anomaly.    We will show that modular covariance is in conflict with the hypothesis
that at enhanced symmetry points the group $F(\Gamma)$ is an automorphism group of the CFT $\CC_{\Gamma}$.

The simplest example of a conflict between  \eqref{eq:NaiveAct}  and modular covariance appears, once again,
in the Gaussian model at the self-dual radius.
Recall that a single compact boson has a Lagrangian specified by
the radius $r$ of the target space circle and the T-duality group acts on the space of
sigma models $O(1,1;\IR)/O(1) \times O(1) \cong \IR_+$ as $r\to \ell_s^2/r$ where $\ell_s$ is
the string length. The T-duality group is isomorphic to $\IZ_2$
and the self-dual radius is an orbifold point of order $2$ in $\CN$. (In this paper we will henceforth
take $\ell_s=1$ so the self-dual radius is $r=1$.) This does \underline{not} imply that there is an action of the
T-duality group on the CFT space associated to the self-dual radius. In fact, only a two-fold covering
group acts on the  CFT space and the only action of T-duality on this state space
consistent with modular invariance is order four.
We will explain these statements in detail in section \ref{sec:GaussianModels}.

\subsection{Doomed To Fail}

The phenomenon we have just described at the points $[\Gamma(\fg)] \in \CN$
arises more generally at the loci where $F(\Gamma)$ is larger than $\IZ_2$.
It is therefore useful to find a criterion for when \eqref{eq:NaiveAct} must be modified, or to put it colloquially,
when implementing \eqref{eq:NaiveAct} in a naive way is ``doomed to fail."  That is, we would like
to know when this naive action of $F(\Gamma)$  on the state space is inconsistent
and we must choose a nontrivial lift $\widetilde{F(\Gamma)}$
to act on the state space (or change the action \eqref{eq:NaiveAct}).  Moreover, we could ask  whether there is a
canonical lift of $F(\Gamma)$ to $\Aut(\CC_{\Gamma})$. In section \ref{subsec:Involution-ModCov} we will show that the $F(\Gamma)$ action
defined by \eqref{eq:NaiveAct} is indeed inconsistent with modular covariance  when there is a nontrivial involution,
\footnote{
We will refer to involutions in $F(\Gamma)$
that are not of the form $p \to - p$ as \emph{nontrivial involutions}. }
say  $g$,
such that there exists a vector $p\in \Gamma$ with $p\cdot g \cdot p$ an odd integer.  More generally, as shown
in section \ref{subsec:ArbEvenOrder},  there is an inconsistency with
modular covariance when there are elements $g\in F(\Gamma)$ of even order $\ell$ such that:
\be\label{eq:LNSV-Cond}
\exists p \in \Gamma \qquad s.t. \qquad
p \cdot g^{\ell/2} \cdot p = 1~ \mod~ 2
\ee
(Of course, $g^{\ell/2}$ is an involution in $F(\Gamma)$, so the problem can always be traced to involutions.)
The criterion \eqref{eq:LNSV-Cond} implies that the subgroup $\langle g\rangle \subset F(\Gamma)$
cannot be lifted to an isomorphic subgroup of $\Aut(\CC_{\Gamma})$ inducing the action of
$\langle g \rangle$ on $\Gamma$ and satisfying \eqref{eq:NaiveAct}.
 We hasten to add that the condition \eqref{eq:LNSV-Cond} does not rule out the existence of some lift
$\widetilde{F(\Gamma)} \subset \Aut(\CC_{\Gamma})$ isomorphic to $F(\Gamma)$. As we will show in
section \ref{subsec:su3-example} in the
explicit example of the $SU(3)$ level one WZW model, it is possible to modify the generators of the
Tits lift by shift vectors so that there is a lift of $W(\fg)_L \times W(\fg)_R$ isomorphic to
$W(\fg)_L \times W(\fg)_R$. What gives is that it is no longer true that $\hat g \vert p \rangle = \vert p \rangle$
where $p$ is in the invariant lattice and  $\hat g\in \Aut(\CC_{\Gamma})$ is a lift of $g$.
%
%
To summarize: The  meaning of the criterion \eqref{eq:LNSV-Cond} is that either:

\begin{enumerate}

\item Equation \eqref{eq:NaiveAct} does not hold for some $p\in \Gamma^g$, or

\item Equation \eqref{eq:NaiveAct} does hold, but $\langle g \rangle \subset F(\Gamma)$ is
lifted to an extension in $\Aut(\CC_\Gamma)$.

\end{enumerate}

We further conjecture that there is in fact a canonical lift of $F(\Gamma)$ to
\be \label{canlift}
\widetilde{F(\Gamma)}^{\rm can} \subset \Aut(\CC_{\Gamma}) \, ,
\ee
given by \eqref{eq:GenRefConj-1}\eqref{eq:GenRefConj-2}. It satisfies the properties  that there is a lifting
$\hat g \in \widetilde{F(\Gamma)}^{\rm can}$ of $g\in F(\Gamma)$ such that
\be\label{eq:CanLift1}
\hat g \vert p \rangle = \vert p \rangle \qquad \qquad  \forall p \in \Gamma^g
\ee
where $\Gamma^g:= \{ p \in \Gamma \vert g\cdot p = p \}$ is the invariant sublattice of $\Gamma$ and
moreover
\be\label{eq:CanLift2}
\hat g^\ell \vert p \rangle =e^{\I \pi p \cdot g^{\ell/2} \cdot p}  \vert p \rangle \qquad \qquad  \forall p \in \Gamma \, .
\ee
As already mentioned, in the
case of CFTs based on $\Gamma(\fg)$ with non-abelian symmetry there is a canonical lift based on  the Tits lift
described in Appendix \ref{sec:WeylLift}.  At the end of section \ref{sec:Involutions} we provide some evidence that
the canonical lift defined by \eqref{eq:GenRefConj-1}\eqref{eq:GenRefConj-2}  is indeed a generalization of the Tits lift.
If the conjecture made in section  \ref{sec:Involutions} is correct then lifting to $\widetilde{F(\Gamma)}^{\rm can}$ at most
doubles the order of any element $g \in F(\Gamma)$.

\subsection{On T-Duality As A Target Space Gauge Symmetry}

The considerations of this paper have some interesting implications for the
relation of the T-duality group to the gauge symmetries of string theory.
Put briefly, it is believed
%
%
that the symmetry groups $F(\Gamma(\fg)) \subset O(d,d;\IZ)$ generate
all of $O(d,d;\IZ)$ except for a $\IZ_2$ transformation that corresponds to world-sheet parity. It is standard string-theory lore \cite{Dine:1989vu,Giveon:1994fu} that
$F(\Gamma(\fg))$ is a subgroup of the  target space $G_L \times G_R$ gauge
symmetry of the target space theory, and therefore $O(d,d;\IZ)$ is a
gauge symmetry of string theory. Unfortunately, this is based on the misconception
that $W(\fg)$ is canonically a subgroup of $G$. Rather, there are subgroups
\be
\widetilde{F(\Gamma(\fg))} \subset \Aut( \CC(\fg))
\ee
lifting $W(\fg)_L \times W(\fg)_R$. These do not fit (in any
way obvious to us) as subgroups of a single common group and hence it is not clear
what, if anything, the different groups $\widetilde{F(\Gamma(\fg))}$ generate.
The main, open, issue can be phrased as follows.

The subgroup of $\CT$ fixing a point $E \in \fB$ that projects to $[\Gamma]$ is isomorphic to $F(\Gamma)$.
As we have just discussed at length, sometimes the   group $F(\Gamma)$ does not lift
to act on the fiber $\CH_E$ over $E$. Only a covering group $\widetilde{F(\Gamma)}$
lifts. Thus, the bundle of CFT state spaces $\pi: \CH \to \CB$ defined above
 does \underline{not} admit the structure of an
$O(d,d;\IZ)$-equivariant bundle. This leaves us with two logical possibilities:

\begin{enumerate}

\item There is a group $\widetilde{\CT}$ acting on $\CH$,  covering the $\CT$ action
on $\CB$, and inducing $\widetilde{F(\Gamma)}$ on the enhanced symmetry locus. Following
the logic of \cite{Dine:1989vu,Giveon:1994fu} it would actually be the group $\widetilde{\CT}$,
rather than $\CT$, which would be a gauge symmetry of string theory.

\item There is no such group $\widetilde{\CT}$. This is a reasonable possibility. Similar phenomena are
quite standard in the study of twisted equivariant K-theory. If this is the case, the idea that
``T-duality is a gauge symmetry of string theory'' is in fact quite mistaken.

\end{enumerate}

Which of the two possibilities is in fact the case is a very interesting question we leave to the future.
The proper resolution of this question will involve an investigation into the moduli \underline{stack}
of toroidal CFTs. Moreover, one must take into account the existence of $U(1)^d \times U(1)^d$ automorphisms of the fiber, i.e. the possibility of combining the transformation with separate left and right $U(1)^d$
automorphisms. These left- and right- $U(1)^d$ automorphisms are also often represented
by asymmetric shift vectors. They act trivially on the base.
We thank D. Freed, D. Freidan, A. Tripathy, and G. Segal for useful discussions about this question.

\subsection{Consistency Conditions For Orbifolds}

Finally, we note that the considerations of this paper are very relevant to orbifold constructions,
namely the gauging of discrete subgroups of the automorphism group of a CFT.  It is important
to bear in mind that the orbifold group is a subgroup of $\Aut(\CC_{\Gamma})$ and is not
a subgroup of $F(\Gamma)$, although much of the literature refers to the orbifold group as
a subgroup of $F(\Gamma)$. In particular, we
note that the criterion \eqref{eq:LNSV-Cond} is closely related to the work of Lepowsky
\cite{LepowskyCalculus,Lepowsky2} as well as to the work of Narain, Sarmadi, and Vafa
\cite{Narain:1986qm} (see their equation $(2.6)$). The work of Lepowsky  addresses a slightly
different problem from that addressed here in that it is concerned with strictly chiral twisted
affine Lie algebras and their modules. Our interpretation of
\eqref{eq:LNSV-Cond} differs from \cite{Narain:1986qm}, where it is
suggested that the condition is a consistency condition in a sense similar to the
level-matching constraints.
We suggest instead the the correct interpretation is as stated above \eqref{canlift}
and that one should only  attempt to construct an orbifold by a subgroup of the lift of $F(\Gamma)$ . This
is consistent with the remarks above equation $(3.3)$ of  \cite{NSVII}.

The consistency conditions for constructing orbifolds have been discussed by a number of
authors \cite{Dixon:1985jw,Dixon:1986jc,Freed:1987qk,Narain:1986qm,Vafa:1986wx}.
A good example is ``level-matching.'' This is an anomaly cancellation condition that is closely
related to modular covariance \cite{Dixon:1986jc,Vafa:1986wx}. The basic point is that the
twisted partition functions $Z(g_t,g_s;\tau)$ described near equation \eqref{eq:Mod-Cov} have
a Hamiltonian interpretation. Namely, there is a space of twisted states $\CH_{g_s}$ (a module
for a twisted vertex operator algebra) and, for $g_t$ in the centralizer of $g_s$, an action
of $g_t$ on $\CH_{g_s}$. Then
\be\label{eq:HamInt}
Z(g_t,g_s;\tau) = \Tr_{\CH_{g_s}} g_t q^{H} \bar q^{\tilde H}
\ee
where $H= L_0 - c/24$ and $\tilde H = \tilde L_0 - \tilde c/24$. The partition function in
the sector of the orbifold theory twisted by $g_s$ is then
\be\label{eq:TwistSecPart}
 \frac{1}{\vert Z(g_s) \vert} \sum_{g_t \in Z(g_s)} \Tr_{\CH_{g_s}} g_t q^{H} \bar q^{\tilde H} \, .
\ee
Of course, $\langle g_s \rangle \subset  Z(g_s)$ so if $g_{t,i}$ is any set of coset representatives for this subgroup
then   \eqref{eq:TwistSecPart} can be written as
\be
\frac{1}{\vert Z(g_s) \vert \ell } \sum_{g_{t,i} } \sum_{k=1}^\ell \Tr_{\CH_{g_s}} g_{t,i} g_s^k  q^{H} \bar q^{\tilde H}
\ee
where the sum on $k$ runs from $1$ to $\ell$, the order of $g_s$. But now
\be
 \sum_{k=1}^\ell \Tr_{\CH_{g_s}} g_{t,i} g_s^k  q^{H} \bar q^{\tilde H} =  \sum_{k=1}^\ell  Z(g_{t,i} ,g_s;\tau-k)
\ee
These averages will all vanish iff there is a modular anomaly in the untwisted sector for some congruence
subgroup of $PSL(2,\IZ)$. The only way to have one of the averages be nonzero is for the   spectrum
of $H-\tilde H$ in $\CH_{g_s}$ to contain an
infinite number of integers. This is the level matching condition.

While level-matching is very powerful one should bear in mind that there can be other consistency conditions.
Indeed the full set of consistency conditions for orbifolds is actually
not known.
\footnote{One could imagine that modular invariance
at higher genus involves new requirements, and this might be the
case for non-abelian orbifold groups. However, in the abelian case
it was shown that no new consistency conditions arise from anomaly
cancellation at higher genus \cite{Freed:1987qk}.}
Clearly, one necessary condition is that the one-loop partition
function of the orbifold theory should have a ``good $q$-expansion.''
This means that $Z$ has a convergent  expansion of the form
\be
Z = \sum_{\mu, \tilde\mu} D_{\mu,\tilde\mu} q^{\mu} \bar q^{\tilde \mu}
\ee
which is not only modular invariant but moreover all the expansion coefficients $D_{\mu,\tilde \mu}$
are nonnegative integers.
\footnote{The $\mu,\tilde \mu$ are arbitrary real numbers in general. The branch of the logarithm
is defined by $q^\mu := \exp[  2\pi \I \mu\tau]$. It is important to note that this
 is true in the bosonic string, which contains no fermions. In superstring theories this
must be modified to account for minus signs due to the presence of spacetime fermion fields.
Nevertheless, one can impose the condition of a good $q$-expansion in the NS sector.}
Moreover  the vacuum has degeneracy one, i.e. the coefficient
of $q^{-c/24} \bar q^{-\tilde c/24}$ must be exactly one.  Of course, given a consistent VOA acting
on a unitary module $Z = \Tr_{\CH} q^{L_0-c/24} \bar q^{\tilde L_0 - \tilde c/24}$ will automatically
have a good $q$-expansion, but in our constructions we  often fall short of defining the full VOA action
on the twisted sectors, so the condition of having a good one-loop $q$-expansion is a useful one.

As we have just mentioned, we believe that \eqref{eq:LNSV-Cond} should not be interpreted
as saying that the CFT orbifold is inconsistent, but rather that there is a nontrivial
lift of the subgroup of $F(\Gamma)$ acting on the Narain lattice to the group of automorphisms
of the CFT $\CC_{\Gamma}$.
In order to support our thesis we demonstrate in section \ref{sec:GaussianModels} that orbifolding by the
$\IZ_4$ group of diagonal T-duality acting on $d$ copies of the Gaussian model at the
self-dual radius satisfies all known consistency conditions, so long as $d=0 ~ \mod ~ 4$.
Similar remarks apply to chiral Weyl reflection orbifolds of the level one $SU(3)$ WZW model.
In fact, given an involution in $F(\Gamma)$ satisfying some conditions stated at the
beginning of   section \ref{sec:GenPF} we show that one can use the method of modular orbits to
 construct a one-loop partition function with a good $q$-expansion
 for the orbifold by $\langle \hat g \rangle \cong \IZ_4$
provided that the associated \emph{twisted characteristic vector}  satisfies
\be
W_g^2 = 0~ \mod ~ 4.
\ee
A twisted characteristic vector is   a vector $W_g \in \Gamma^g$ such that
\be\label{eq:TwistCV}
p \cdot g^{\ell/2} \cdot p = W_g \cdot p~  \mod ~ 2 \qquad \qquad \forall p \in \Gamma ~ ,
\ee
where $\ell$ is the (even) order of $g$.
The vector $W_g$ is only defined modulo $2 \Gamma$ and generalizes the notion of a characteristic vector
of an odd lattice. For more details see section \ref{sec:Involutions}.
\footnote{ The vector $W_g  ~\mod ~ 2 \Gamma$
should have a topological interpretation in terms of the $G$-equivariant $E^4$-cohomology of $BT$ for a suitable torus $T$,
where $G = \langle g \rangle$. This
 interpretation should play a role when interpreting our results in terms of three-dimensional
Chern-Simons theory. We leave such considerations to the future.}

\subsection{Future Directions}

The above discussion begs the question: What are the consistency conditions for toroidal orbifolds?
It is possible that the application of recent ideas relevant to the classification of
symmetry-protected topological phases of matter can be usefully applied to this problem.
We have had some initial discussions about this idea with  D.~Gaiotto and N.~Seiberg
and we hope to develop this approach further in the future. Moreover, one
 can interpret many aspects of our discussion  in the language of defects
\cite{Fuchs:2007tx} and it might be fruitful to use the language of defects to approach
the more general question of consistency conditions for asymmetric orbifolds.

Finally we discuss some possible consequences of our results. As mentioned earlier, the reinterpretation
of \eqref{eq:LNSV-Cond} presented here and in \cite{NSVII} allows for a more general class of asymmetric orbifold constructions. We were in fact led to
the considerations of this paper precisely by the study of such constructions in the context of work on moonshine
and string duality which will appear in \cite{css}.  We expect that there will be additional consequences for the study
of moonshine.
%
For example one might wonder if there are interesting consequences
for the ``symmetry surfing" proposal of \cite{symsurf,Taormina:2013mda,Gaberdiel:2016iyz}.
We hope to explore some of these potential consequences in future work.

\section{Products Of Self-Dual Gaussian Models }\label{sec:GaussianModels}

We now use the Gaussian model at the self-dual radius as a simple model to diagnose the structure of T-duality,
the conditions following from modular covariance, and the construction of asymmetric orbifolds by T-duality.
\footnote{This section has considerable overlap with
section four of \cite{Aoki:2004sm}.}
We first consider
a single Gaussian model and then in order to construct consistent asymmetric orbifolds, $d$ copies of the Gaussian model.

The $c=1$ Gaussian model (see \cite{Ginsparg:1988ui} for a review) is described by a single real bosonic field $X$
with action
\be
S= \frac{r^2}{4\pi\ell_s^2} \int d\tau \int_0^{2\pi} d\sigma \left[ (\p_\tau X)^2 - (\p_\sigma X)^2\right]
\ee
with periodicity   $X \sim X+ 2 \pi $. In the context of string theory it describes string propagation on a target space circle of radius
$r$.
The momentum and winding zero modes of the Gaussian field are defined by the general solution of the equation of motion:
\be
X = x_0 + \frac{p_L}{\sqrt{2}}(\tau + \sigma) + \frac{p_R}{\sqrt{2}} (\tau-\sigma) + X^{\rm osc}
\ee
where we have set $\ell_s=1$ and $X^{\rm osc}$ is the sum of solutions with nonzero Fourier modes. The zero modes
have the property that the vectors $(p_L,p_R)$ are valued in
an even unimodular lattice embedded in $\IR^{1;1}$. The lattice of zero modes can be written as
\be
\Gamma(r) := \{ n e_r + w f_r \vert n,w\in \IZ \} \subset \IR^{1;1}
\ee
where
\be
e_r =\frac{1}{\sqrt{2}} (1/r;1/r), \qquad f_r=\frac{1}{\sqrt{2}} (r;-r)
\ee
Note that $e_r^2=f_r^2=0$, $e_r\cdot f_r=1$ so that $\Gamma(r)$ is indeed an embedding of the even unimodular
(a.k.a. self-dual Lorentzian) lattice $II^{1,1}$ of rank $2$ and signature $(1,1)$. Note that
the CFT is invariant under $O(1)_L \times O(1)_R \cong \IZ_2 \times \IZ_2$.  Choose generators
of this automorphism group:
\be
\begin{split}
\sigma_L  : (X_L, X_R) &  \rightarrow (-X_L, X_R) \\
\sigma_R : (X_L, X_R) &  \rightarrow (X_L, - X_R) \\
\end{split}
\ee
Then
\be
\sigma_L\cdot \Gamma(r) = \sigma_R \cdot \Gamma(r) = \Gamma(1/r)
\ee
This proves that the moduli space $\CN$ of CFTs is related to the
 space of sigma models $O(1,1;\IR)/O(1)\times O(1) \cong \IR_+$, parametrized by $r$, by the
quotient by $r \to 1/r$.  Note that   $F(\Gamma(r))\cong \IZ_2$ for $r\not=1$ and $F(\Gamma(r))\cong \IZ_2 \times \IZ_2$
at the self-dual radius $r=1$. This is the $\IZ_2$ orbifold point of the Narain moduli space
$\CN \cong \IR_+/\IZ_2  \cong [1,\infty)$. In this case the enhanced symmetry locus is a single
(orbifold) point. We can say that $\sigma_L$ and $\sigma_R$ are left- and right- moving
T-duality symmetries. Note that, with our particular choice of basis for $\Gamma(r=1)$ the
automorphism $\sigma_R$ is just
\be
e \leftrightarrow f
\ee
and we will focus on this T-duality below. Here and henceforth we simply
denote $e_r, f_r$ at $r=1$ by $e,f$.  Of course $\sigma_L \sigma_R = -1$ is the trivial involution.
  Note that if we identify the positive root of $\fs\fu(2)$ with
$\sqrt{2}\in \IR$ and the dominant fundamental weight with $1/\sqrt{2}$ then we can identify $\Gamma(r=1)$
with $\Gamma(\fs\fu(2))$ defined in equation \eqref{eq:ESP-DEF} above.

The easiest way to see that there is an order four action lifting the T-duality action is to consider the
$\fs\fu(2)_L \oplus \fs\fu(2)_R$ current algebra symmetry of the Gaussian model at the self-dual point.
The left- and right-moving currents are
\be
J^3(z) = \frac{1}{\sqrt{2}} \p X_L(z) \qquad  J^\pm(z) = : e^{\pm \I \sqrt{2} X_L(z)} :\hat c
\ee
\be
\tilde J^3(\bar z) = \frac{1}{\sqrt{2}} \bar\p X_R(\bar z) \qquad  \tilde J^\pm(\bar z) = : e^{\pm \I \sqrt{2} X_R(\bar z)} :\hat c
\ee
where the tilde indicates right-moving symmetry and $\hat c$ is a cocycle factor discussed below. The T-duality transformation
leaves the left-moving currents unchanged but takes $\tilde J^3 \to - \tilde J^3$ and $\tilde J^\pm \to \tilde J^\mp$.
It therefore acts as a $180$-degree rotation on the Lie algebra $\fs\fu(2)_R$. On the other hand, the states with
$(n=0, w=\pm 1)$ and $(w=0, n=\pm 1)$ transform in the $2_L \otimes 2_R$ of the $SU(2)_L \times SU(2)_R$ global symmetry.
Hence, to define the action on the Hilbert space we must lift T-duality to an order four action.
Thus, despite appearances, \emph{T-duality of the Gaussian model at the self-dual point is order four!}
\footnote{A similar surprise was noted by W. Nahm and K. Wendland concerning mirror symmetry
of Kummer surfaces in \cite{Nahm:2001kh}. While similar in spirit the two remarks are different.
In the example of mirror symmetry, the observation is that the action on the sigma model moduli
space is order four. Here, the action on the sigma model moduli space is order two, but its action
on the CFT state space is order four.}
We will generalize this discussion in section \ref{sec:NonabelianSymmetry} below.

As an aside we note that while  the $SU(2)$ level one WZW model has $SU(2)_L \times SU(2)_R$, the diagonally embedded center generated by
$(-1,-1)$ acts ineffectively so in fact the symmetry is $SU(2)_L \times SU(2)_R/\IZ_2$. The lift of the full
enhanced symmetry group $F(\Gamma(r=1))$ is then $(\IZ_4 \times \IZ_4)/\IZ_2$.

Now we turn to the modular covariance approach.  The Hilbert space
of states has sectors labelled by $n,w$ and  each sector consists of the usual Fock space of states formed by acting on the
vacuum with creation operators $\alpha_{-n}$,$\tilde\alpha_{-n}$, $n \in \IZ$. The modular invariant partition function is given by
\be
Z(\tau)= B_+ \overline{B_+} \Theta_{\Gamma} = \frac{ \sum_{(p_L, p_R) \in \Gamma^{1,1}} q^{p_L^2/2} {\bar q}^{p_R^2/2}}{ \eta \bar \eta}
\ee
where $\eta(\tau) = q^{1/24} \prod_{n=1}^\infty(1-q^n)$ with $q=e^{2 \pi i \tau}$ is the Dedekind eta function. We have
also introduced the notation
\be\label{eq:BPlusMinus-def}
B_\pm := \frac{1}{q^{1/24} \prod_{n=1}^\infty (1 \mp q^n)}
\ee
so that $B_+(\tau) = 1/\eta(\tau)$ and $B_-(\tau)$ is the trace of $-1$ acting on the chiral
oscillators:
\be\label{eq:Bplus-def}
B_- = \frac{\eta(\tau)}{\eta(2\tau)} = \frac{\vartheta_4(2\tau)}{\eta(\tau)}.
\ee
and will be useful presently.
Finally, $\Theta_{\Gamma}$ is the Siegel-Narain theta function. Our conventions for theta functions are spelled out
in Appendix \ref{app:ThetaFunctions}. In terms of modular functions the untwisted torus partition function can be
written as
\be\label{eq:Z11}
Z(1,1) = \frac{1}{\eta\bar\eta}  \left( \vartheta_3(2\tau)\overline{\vartheta_3(2\tau) }  + \vartheta_2(2\tau)
\overline{\vartheta_2(2\tau) } \right)
\ee
and this turns out to be modular invariant.

Now let us assume (counter-factually, as we have just seen using $SU(2)$ invariance) that there is a lift of $\langle \sigma_R\rangle$
to a $\IZ_2$ group of automorphisms acting on the CFT. Let $\hat g$ be the generator of this purported lift.
\footnote{We are deprecating the notation $U( g)$ used in section \ref{subsec:TechSumm} in favor
of $\hat g$ for simplicity.}
To evaluate $Z(\hat g, 1)$ use the relation between the one-loop partition function and a trace on the Hilbert
space. The naive action on the Hilbert space is defined by choosing a basis
\be
\vert A, \tilde A; p \rangle
\ee
where $A,\tilde A$ is shorthand for oscillator states, so that
\be
\hat g \vert A, \tilde A; p \rangle = (-1)^{\tilde N}
\vert A, \tilde A; g\cdot p \rangle
\ee
where $p= ne + w f $ and $\tilde N$ is the number of right-moving oscillators in the state .
 Invariance of the momentum forces $p_R=0$, or equivalently $n=w$ so the momentum is purely
left-moving. The phase then simplifies to $(-1)^{\tilde N}$ coming
from the straightforward action on the oscillators.  The resulting partition function is
\be\label{eq:Z10}
Z(\hat g, 1) = \frac{ \vartheta_{3}(2\tau) \overline{\vartheta}_4(2\tau)  }{\eta(\tau) \bar \eta(\tau)} \, .
\ee
Modular covariance now forces
\be
Z(1,\hat g) = 2^{-1} \frac{1}{\eta(\tau) \bar \eta(\tau)}  \vartheta_{3}\left( \tau/2 \right) \overline{\vartheta}_2 \left( \tau/2 \right)
\ee
up to a phase, and then again using modular covariance we must have
\be\label{eq:Z-gsectors}
\begin{split}
Z(1,\hat g) &= 2^{-1} \frac{1}{\eta(\tau) \bar \eta(\tau)}  \vartheta_{3} \left( \tau/2 \right) \overline{\vartheta}_2 \left( \tau/2 \right)  \\
Z(\hat g,\hat g) &= 2^{-1} \frac{1}{\eta(\tau) \bar \eta(\tau)}  \vartheta_{3} \left( (\tau-1)/2 \right) \overline{\vartheta}_2 \left( (\tau-1)/2 \right)  \\
Z(\hat g^2,\hat g) &= e^{\I \pi/4} 2^{-1} \frac{1}{\eta(\tau) \bar \eta(\tau)}  \vartheta_{4} \left( \tau/2 \right) \overline{\vartheta}_2 \left( \tau/2 \right)  \\
Z(\hat g^3,\hat g) &= e^{\I \pi/4} 2^{-1} \frac{1}{\eta(\tau) \bar \eta(\tau)}  \vartheta_{4} \left( (\tau-1)/2 \right) \overline{\vartheta}_2 \left( (\tau-1)/2 \right)  \\
Z(\hat g^4,\hat g) & = e^{\I \pi/2} 2^{-1} \frac{1}{\eta(\tau) \bar \eta(\tau)}  \vartheta_{3} \left( \tau/2 \right) \overline{\vartheta}_2 \left( \tau/2 \right)  \\
\end{split}
\ee

Note, particularly, that $Z(\hat g^2, \hat g)$ is not proportional to $Z(1,\hat g)$. Thus, there cannot be an order two
action on the CFT space. It is true, however, that $Z(\hat g^4, \hat g)$ is proportional to $Z(1, \hat g)$ suggesting that
T-duality might lift to an order four action on the CFT space. This will prove to be correct.

Indeed we can define an   action of an order four lift of   $\sigma_R$ as follows:
\be
\hat g  \vert A, \tilde A; p \rangle = (-1)^{\tilde N} e^{ \frac{\I \pi}{2} (n+w)^2  }
\vert A, \tilde A; g\cdot p \rangle \, .
\ee
This can be derived using the discussion of
cocycles in section \ref{sec:CocycleComments} below. In particular this is order four. Note that
\be
\hat g^2 \vert p \rangle = e^{\I \pi p\cdot g\cdot p} \vert p \rangle \qquad\qquad \forall p \in \Gamma(\fs\fu(2))
\ee
in accord with our conjecture \eqref{eq:CanLift2}.

Now let us turn to the orbifold by T-duality, or rather, by the group (isomorphic to $\IZ_4$) generated by
$\hat g$. It follows immediately from \eqref{eq:Z-gsectors} that there is a modular anomaly in the
$\hat g$-twisted sector (or equivalently, an anomaly under $ST^4S$ in the untwisted sector). Therefore
there is no consistent T-duality orbifold of the Gaussian model, assertions to the contrary in the
literature notwithstanding. Equivalently, level matching fails for this model.

We could, however, consider a product of $d$ copies of the Gaussian model
at the self-dual radius and consider the orbifold by the simultaneous $r\to 1/r$ duality on all the circles.
That is, we are considering the point in the moduli space of $d$ bosons on a torus with
$\fg = \fs\fu(2)^{\oplus d}$. Concretely
\be
\Gamma(A_1^d) = \{ (r + \lambda; r' + \lambda) \}
\ee
where $r,r'$ are root vectors  and
\be
\lambda = \half \sum_i \epsilon_i \alpha_i \qquad \epsilon_i \in \{ 0, 1\}
\ee
where $\alpha_i$ is the simple root for the $i^{th}$ summand. We let $g$ be the lattice
automorphism taking $(p_L; p_R ) \rightarrow (p_L ; - p_R)$. Once again the lifted
action $\hat g$ on the CFT defined by the diagonal action of the
lift of T-duality on a single Gaussian model will be order four. We can try to
orbifold by the $\IZ_4$ group $\langle \hat g \rangle$. This is of some interest since the alleged
inconsistency condition \eqref{eq:LNSV-Cond} is met for such models for all values of $d$.
To see this note that it is met by choosing all but one of the $\epsilon_i$ to vanish.

Let us examine first the level matching condition.
A single $\IZ_2$-twisted boson has a ground state energy $+\frac{1}{16}$. Therefore
in the twisted sector a state has quantum numbers $N_L$, the left moving oscillator
level (this is an integer) and $N_R$ the right-moving oscillator number (this is in $\half\IZ$
since  the right-moving oscillators are half-integer moded: $\tilde \alpha_{-r }$, $r\in \IZ+\half$).
In addition there is a momentum in the dual of the invariant sublattice: $p \in (\Gamma^g)^\vee$.
For very general reasons explained below $2p^2$ is an integer, so $\half p^2 \in \frac{1}{4}\IZ$.
In our specific example
\be
(\Gamma^g)^\vee= \left\{ \left( \sum_i n_i \frac{\alpha_i}{2};0 \right)   \vert n_i \in \IZ \right\} \, .
\ee
Now, because the twisted sector ground state of a single real $\IZ_2$-twisted boson
has energy $1/16$, a state with
quantum numbers $(N_L, N_R, p)$ will have equal left and right scaling dimensions if:
\be\label{eq:EqualEnergy}
 \left( N_L  +\half p_L^2 - \frac{d}{16} - N_R -\half p_R^2 \right) = 0
\ee
Since $N_L$ can be an arbitrary nonnegative integer and $N_R\in \half \IZ $
this becomes a condition on $p$ and $d$. That condition is
\be\label{eq:LevelMatching}
 p^2 - \frac{d}{8} = 0~ \mod ~1
\ee
Now, since $p^2\in \half \IZ$ we see that this condition can only be satisfied for $d=0~ \mod ~4$.

We claim that all known consistency conditions are satisfied for the $\IZ_4$ orbifold for $d=0~\mod ~4$.
We have just checked level matching. In addition to level-matching, one should
check that the partition function has a good $q$-expansion in the sense explained in
section \ref{subsec:TechSumm}.

We can easily compute the partition function for the orbifold of $A_1^d$
using modular covariance and the partition functions computed above for the $A_1$ theory since
\be
Z_{A_1^d}(g,h) = ( Z_{A_1}(g,h))^d \, .
\ee
The only tricky point is the $\hat g^2$-sector. Using modular covariance one easily computes
\be
\begin{split}
Z(\CH_{\hat g^2}) & = \frac{1}{4}  \frac{1}{(\eta\bar\eta)^{4k}} [ \left( \vartheta_3(2\tau) \bar\vartheta_2(2\tau)+ \vartheta_2(2\tau)\bar\vartheta_3(2\tau)\right)^{4k}\\
 & +   \left( \vartheta_3(2\tau) \bar\vartheta_2(2\tau)- \vartheta_2(2\tau)\bar\vartheta_3(2\tau)\right)^{4k} + 2(-1)^k \vartheta_2(2\tau)^{4k} \bar\vartheta_4(2\tau)^{4k} ] \\
\end{split}
\ee
where $d=4k$. The second line contains contributions with minus signs which are potentially problematic.
However,   the terms in square brackets   can be written as:
\be
\begin{split}
2 \sum_{s=0}^{2k-1} {4k  \choose 2s} (\vartheta_3\bar \vartheta_2)^{4k-2s} (\vartheta_2 \bar \vartheta_3)^{2s}
& + 2 \vartheta_2^{4k} \left( \bar \vartheta_3^{4k} + (-1)^k \bar \vartheta_4^{4k} \right) \\
\end{split}
\ee
The first sum manifestly has a good $q$-expansion. The only possibly problematic part
is the second term. For $k=1$ we note that positivity of this term follows from Jacobi's abstruse identity,
$\vartheta_3^4-\vartheta_4^4= \vartheta_2^4$.  For general $k$ we write this term as
\be
\left( \bar \vartheta_3^{4k} + (-1)^k \bar \vartheta_4^{4k} \right) =
\sum_{n_1, \dots, n_{4k}} \bar q^{\half (n_1^2 + \cdots + n_{4k}^2)} (1 + (-1)^{k+n_1 + \cdots + n_{4k} } )
\ee
and note that the coefficients of $\bar q^{\ell/2}$ in this expression are either $0$ or $2$.
The entire expression in square brackets is of the form $2^{1+4k}$ times a good
$q$-expansion and hence $Z(\CH_{\hat g^2})$ has a good $q$-expansion.
One can similarly check that $Z(\CH_{\hat g})$ and $Z(\CH_{\hat g^3})$ have good $q$-expansions.

\section{Models With Non-Abelian Symmetry}\label{sec:NonabelianSymmetry}

Let $\fg$ be a semi-simple (but not necessarily simple) and simply-laced Lie algebra of full rank.
The points $\Gamma(\fg)$ of the Narain lattice defined in \eqref{eq:ESP-DEF} are very special.
%
%
 The CFT $\CC(\fg)$ corresponding to these points  is isomorphic to the
 WZW model at level one for the simply connected covering group $G$.
When $\fg$ is simple the CFT space of the WZW model is
\be\label{eq:ESP-StateSpace}
\CH = \oplus_{\theta\cdot \lambda \leq 1}  V_{\lambda} \otimes \overline{V_{\lambda}}
\ee
and is a representation of $\widetilde{LG}_L \times \widetilde{LG}_R $ although the diagonally embedded
center of $G$ acts trivially. Here $\theta$ is the highest root and $V_{\lambda}$ is the
integrable lowest weight representation. In particular, the subgroup of constant loops  $G_L \times G_R$ acts.
On the other hand, in the equivalent formulation in terms of free bosons on a torus,
the crystallographic symmetry group given by \eqref{eq:ESP-Crystal} acts canonically
on the oscillators and momenta of the theory. Nevertheless, as we have repeatedly stressed, this group must
\underline{not} be confused with a group of automorphisms of the CFT $\CC(\fg):=\CC_{\Gamma(\fg)}$. In particular,
there is no natural action of it on the state space \eqref{eq:ESP-StateSpace} compatible with
the action on the oscillators and momenta. We now discuss this in a little more detail.

First we compute $F(\Gamma(\fg))$. As is well-known, the
 automorphism group $\Aut(\Lambda_{wt}(\fg))$ is the semidirect product
 $W(\fg) \rtimes \CD(\fg)$ where $\CD(\fg)$ is the group of outer automorphisms
 of $\fg$ \cite{FultonHarris,Humphreys}.
The group $F(\Gamma(\fg))$ is thus the semidirect product
\be
F(\Gamma(\fg)) = \left( W(\fg)_L \times W(\fg)_R \right) \rtimes \CD(\fg)
\ee
where $\CD(\fg)$ acts diagonally on  $W(\fg)_L \times W(\fg)_R $.
\footnote{It is worth noting that the definition of $\Gamma(\fg)$ can be generalized to
an even unimodular lattice $\Gamma(\fg,\sigma)$ defined by any element $\sigma\in \Aut(\fg)$
by choosing pairs $(p_L;p_R) \in \Lambda_{wt}(\fg) \times \Lambda_{wt}(\fg) $ such that
$p_L - \sigma(p_R) \in \Lambda_{rt}(\fg)$. These lattices project to the same point in $\CN$. }

Let us begin by considering the lift of the subgroup $ W(\fg)_L \times W(\fg)_R$. Recall
the discussion around \eqref{eq:WG-Def} and \eqref{eq:WG-Seq} of section \ref{subsec:TechSumm}.
In order to define an automorphism group of the CFT $\CC(\fg)$ inducing the action of $F(\Gamma(\fg))$
on the oscillators we must lift $W(\fg)$ to a subgroup of $N(T)\subset G$ and use the action defined by
$G\subset \widetilde{LG}$.
That is, we must choose a finite subgroup $\widetilde W(\fg)\subset N(T)$ so that if $\pi: \widetilde W(\fg) \to W(\fg)$
then for every $\tilde g \in \widetilde W(\fg)$ we have $\tilde g t \tilde g^{-1} = \pi(\tilde g)\cdot t$.

We now explain in more detail how subgroups of $N(T)_L \times N(T)_R$ act on the CFT space. We can
choose a basis of states for $\CH$ of the following form. We begin with the representation
\be\label{eq:FinDimRep}
\oplus_{\theta\cdot \lambda \leq 1}  R_{\lambda} \otimes \overline{R_{\lambda}}
\ee
of the finite-dimensional group $G_L \times G_R$. Here $R_{\lambda}$ is
 the irreducible representation of  $G$ with dominant weight $\lambda$.
Now choose a weight basis for \eqref{eq:FinDimRep} and denote it:
\be\label{eq:FKS-basis}
\vert \mu_L \rangle \otimes
\overline{
 \vert \mu_R \rangle } \, .
\ee
Note that $\mu_L, \mu_R$ are weights in the same irreducible representation $R_\lambda$ and hence
 $\mu_L - \mu_R$ is in the root lattice.
 Next we act on this basis with arbitrary monomials of
raising operators for both the left and right-moving current algebra symmetry.
The raising operators are either of the form $\alpha_I \cdot H_{-n}$
where $\alpha_I$ are simple roots and $n>0$ labels the
Fourier modes of the current, or they are of the form $E^{\alpha}_{-n}$ where again $n>0$ labels a Fourier
mode and $\alpha$ is a root. The resulting set of states is an overcomplete set in general (because of null vectors)
but it will suffice to specify the group action on this set.

An element $(\hat g_L, \hat g_R)\in N(T)_L \times N(T)_R \subset G_L \times G_R$ preserves the currents.
For example:
\be\label{eq:ConjCurrent}
\begin{split}
\hat g_L E_{-n}^{\alpha} \hat g_L^{-1} & =  E_{-n}^{\bar g_L \cdot \alpha}\\
\hat g_L \alpha \cdot H_{-n}  \hat g_L^{-1} & = (\bar g_L \cdot \alpha)\cdot H_{-n} \\
\end{split}
\ee
where $\bar g_L \cdot \alpha$ is the induced action of the projection of $\hat g_L$ in $N(T)/T := W$
on the root lattice, and similarly for $\hat g_R$. The action \eqref{eq:ConjCurrent} will map null
vectors to null vectors so to define the action on the states we need only define the action on the states
\eqref{eq:FKS-basis}  and this is:
\be
(\hat g_L, \hat g_R)\cdot \left(    \vert \mu_L \rangle \otimes \overline{   \vert \mu_R \rangle } \right)
:=
  R_\lambda(g_L) \vert \mu_L \rangle \otimes
 \overline{   R_{\lambda}(g_R) \vert \mu_R \rangle }.
\ee
Note that states of the form \eqref{eq:FKS-basis}
correspond to states $\vert p \rangle$ in the vertex operator algebra construction with momentum
\be
p = (  \mu_L ;  \mu_R)
\ee
so together with \eqref{eq:ConjCurrent} we see that $(\hat g_L, \hat g_R)$ acts on the Narain
lattice  through the projection to the Weyl group.

Lifting the Weyl group to a subgroup of $N(T)$ has been studied in the mathematical literature and
we review some relevant results in Appendix \ref{sec:WeylLift} below. The key points are that there is always
a canonical lift $\widetilde W(\fg)^{\rm T} $ called the \emph{Tits lift}, but $\widetilde W(\fg)^{\rm T} $
is never isomorphic to the Weyl group: The lift of reflections in simple roots are elements of order four in $N(T)$.
 For some groups there do exist lifts isomorphic to $W(\fg)$ but for
some groups no such lift exists. It is possible to be quite explicit about the various possibilities, see for example \cite{CWW,dw,adamhe}
and Appendix \ref{app:SUN-Tits}.

\subsection{Example: Products Of $SU(3)$ Level One }\label{subsec:su3-example}

A very useful example is the model $\CC(\fs\fu(3))$ with $g$ a right-moving involution corresponding
to reflection in a simple root. In this case one can modify the generators of the Tits lift by shift vectors so that
there is a lift of $F(\Gamma)$ isomorphic to $F(\Gamma)$, even though the condition \eqref{eq:LNSV-Cond}
is satisfied.

As discussed in Appendix \ref{sec:WeylLift} below, if we take $T$ to be the subgroup of
diagonal $SU(3)$ matrices then lifts of the Weyl reflections in $\alpha_1, \alpha_2$ must have the form
\be\label{eq:Gen-g1}
\hat g_1= \begin{pmatrix} 0 & x_1 & 0  \\   y_1 & 0 & 0 \\  0 & 0 & z_1 \\  \end{pmatrix}
\ee
\be\label{eq:Gen-g2}
\hat g_2 = \begin{pmatrix} z_2 & 0 & 0  \\   0 & 0 & x_2 \\  0 & y_2  & 0 \\  \end{pmatrix}
\ee
where $x_i y_i z_i=-1$. Conjugation on $T$ by these matrices will induce the action of the Weyl reflections
in $\alpha_1, \alpha_2$, where we choose the standard simple roots.
If we choose
\be\label{eq:WeylLift1}
\hat g_1^W = \begin{pmatrix} 0 & 1 & 0 \\   1 & 0 & 0 \\  0 & 0 & -1 \\  \end{pmatrix}
\ee
\be\label{eq:WeylLift2}
\hat g_2^W  = \begin{pmatrix}  -1  & 0  & 0 \\    0 & 0 & 1 \\  0 & 1 & 0 \\  \end{pmatrix}
\ee
then $\hat g_1, \hat g_2 \in SU(3)$ generate a subgroup of $N(T)$ isomorphic to $S_3$. On
the other hand  the Tits lift is
\be
\hat g_1^T = \exp[ \frac{\pi}{2}(e_1-f_1) ] = \begin{pmatrix} 0 & 1 & 0  \\   -1 & 0 & 0 \\  0 & 0 & 1 \\  \end{pmatrix}
\ee
\be
\hat g_2^T = \exp[ \frac{\pi}{2}(e_2-f_2) ] = \begin{pmatrix} 1 & 0 & 0  \\   0 & 0 & 1 \\  0 & -1  & 0 \\  \end{pmatrix} \,
\ee
(where $e_i, f_i$ are Serre generators).
Note that $\hat g_i^T$ are both of order four, so they generate an extension of $S_3$ by $\IZ_2 \times \IZ_2$.
For later use note that if we compare the ``Weyl lift'' \eqref{eq:WeylLift1} and \eqref{eq:WeylLift2} with the
Tits lift then we have
\be
\hat g_1^{W} = \hat g_1^{T} t_1
\ee
with
\be
t_1 = \begin{pmatrix} -1 &  &  \\  & 1 &   \\   & & -1  \\  \end{pmatrix} \, .
\ee
Note that this acts on the weight basis as
\be
t_1 \vert \mu \rangle = e^{\pi \I \theta \cdot \mu} \vert \mu \rangle
\ee
where $\theta=\alpha_1 + \alpha_2$ is the highest root. Similarly, one may check
that $\hat g_2^W = \hat g_2^T t_2$ where $t_2 \vert \mu \rangle =  e^{\I \pi \alpha_1 \cdot \mu } \vert \mu \rangle$
in the three-dimensional defining representation.

Turning now to the CFT $\CC(\fs\fu(3))$ the vectors in the Narain lattice are
of the form
\be\label{eq:gv2}
(n_1 \alpha_1 + n_2 \alpha_2 + r \lambda^2; \tilde n_1 \alpha_1 + \tilde n_2 \alpha_2 + r \lambda^2)
\ee
where $\alpha_i$ are the simple roots and $\lambda^i$ the dual fundamental weights and
$n_i, \tilde n_i \in \IZ$ and $r=0,1,2$. We are going to consider  a
symmetry which acts on the Narain lattice as a right-moving reflection in the simple root $\alpha_1$:
\be
g\cdot ( p_L; p_R ) := (p_L ; \sigma_{\alpha_1}(p_R)) \, .
\ee
The condition \eqref{eq:LNSV-Cond} is satisfied iff $\tilde n_2$ is odd because:
\be
p\cdot g p = p\cdot p + (2\tilde n_1 - \tilde n_2) (\tilde n_1 \alpha_1 + \tilde n_2 \alpha_2 + r \lambda^2) \cdot \alpha_1
= \tilde n_2 ~\mod ~2
\ee
We can choose a twisted characteristic vector  $W_g \in \Gamma^g$ (see equation
\eqref{eq:TwistCV} and section \ref{subsec:Involution-ModCov}) to be
\be
W_g = (0;\alpha_1 + 2\alpha_2)
\ee
so that $p\cdot g p = p \cdot W_g ~\mod~ 2$ for all vectors $p\in \Gamma$.

The action of $\hat g_1^T$ on $\CC(\fs\fu(3))$ satisfies \eqref{eq:CanLift1} and
\eqref{eq:CanLift2} so the  discussion of modular covariance with respect to twisting by this action is
very similar to that for T-duality in the Gaussian model. We have
\be\label{eq:Zg1-su3}
Z(\hat g, 1) = B_+^2 \overline{B_+ B_-} \Theta_{\Gamma^g}(\tau, 0,0)
\ee
\be
Z(\hat g^2, 1) = B_+^2 \overline{B_+^2} \Theta_{\Gamma}(\tau,-\half W_g,0)
\ee
where we recall that $B_\pm$ were defined in \eqref{eq:BPlusMinus-def} and
$\Theta_{\Gamma^g}$ is the theta function of the invariant sublattice under
the action of $g$.

Applying the $S$-transformation to \eqref{eq:Zg1-su3} we get
\be
\begin{split}
Z(1, \hat g)(\tau) &= B_+^2 \overline{B_+ T_-} \frac{1}{2} \Theta_{(\Gamma^g)^\vee}(\tau, 0 ,0 )\\
\end{split}
\ee
where
\be
T_-(\tau) :=  \frac{\vartheta_2(\tau/2)}{\eta(\tau)} \, .
\ee
%
%
%

Using $T_-(\tau+2) = e^{2\pi \I/24} T_-(\tau)$
it is  easy to check that under $\tau \to \tau +2$ this function is not covariant, but
\be
Z(\hat g^{-4}, \hat g)(\tau)  = Z(1, \hat g)(\tau+4)
 = e^{-4\pi\I/8} \frac{1}{2} B_+^2 \overline{B_+^2} \bar\vartheta_2(\tau/2)   \Theta_{(\Gamma^g)^\vee}(\tau, 0 ,0 )
\ee

As in the case of the Gaussian model, we can consider the orbifold of the direct product $\CC(\fs\fu(3))^d$ by the
$\IZ_4$ group generated by the diagonal action of $\hat g_1^T$. Level matching is only satisfied for $d=0 ~\mod ~4$
and with a little patience one can check that the partition function indeed has a good $q$-expansion.

It is interesting to compare the above discussion with the analogous one for the Weyl lift $\hat g_1^W$.
This differs from the Tits lift  by a shift vector $e^{2\pi \I \hat p \cdot s }$ with $s =(0; \half \theta)$
and now we can compute
\be
\begin{split}
(\hat g_1^T) e^{2\pi \I \hat p \cdot s } (\hat g_1^T) e^{2\pi \I \hat p \cdot s }&  =
(\hat g_1^T)^2  e^{\pi \I \hat p \cdot (0; \alpha_2 + \theta)}\\
&  = (\hat g_1^T)^2 e^{\pi \I \hat p \cdot (0; 2\alpha_2 + \alpha_1)} \\
\end{split}
\ee
On the other hand
\be
(\hat g_1^T)^2 = e^{\pi \I \hat p \cdot W_g} = e^{\pi \I \hat p \cdot (0; \alpha_1) }
\ee
and hence $\hat g_1^W = \hat g_1^T e^{2\pi \I \hat p \cdot s }$ has order two acting on $\CC(\fs\fu(3))$.
One can confirm that the partition functions have the correct modular covariance:
\be
Z(1, \hat g_1^W)(\tau+2) = - e^{- 2\pi \I/8} Z(1, \hat g_1^W)(\tau).
\ee
In checking this one must bear in mind that if $p\in \Gamma^g$ is in the invariant lattice then
\be\label{eq:g1W-expl}
\hat g_1^W \vert p \rangle = e^{\I \pi (\tilde n_1 + r)} \vert p\rangle
\ee
in the parametrization used in \eqref{eq:gv2}. (In this parametrization the
invariant lattice is defined by the condition $\tilde n_2 = 2 \tilde n_1$.)
Note that \eqref{eq:g1W-expl} violates \eqref{eq:NaiveAct}, even for $p\in \Gamma^g$.

If we now consider the  orbifold of the direct product $\CC(\fs\fu(3))^d$ by the
$\IZ_2$ group generated by the diagonal action of $\hat g_1^W$ then level-matching -
or, equivalently, the absence of modular anomalies requires   $d=0~\mod ~8$.
Once again, one can check that the partition function of the orbifold theory
has a  good $q$-expansion. Thus, the asymmetric orbifold by the $\IZ_2$ group
generated by $\hat g_1^W$ satisfies all known consistency conditions.

\subsection{A Nontrivial Lift Of An Outer Automorphism Of $\fg$}

Thus far we have discussed involutions in the subgroup $W(\fg)_L \times W(\fg)_R \subset F(\Gamma(\fg))$.
It is also interesting to ask about group elements projecting to nontrivial members of $\CD(\fg)$.
This group is described in \cite{FultonHarris}. For $\fa_n$ the diagram automorphism just corresponds to
complex conjugation on $\fs\fu(n+1)$. It acts as $-1$ on the lattice $\Gamma(\fg)$ and is thus a trivial
involution and the lift is order two. Similarly for $\fd_n$ the group is $\IZ_2$. It corresponds to a parity
transformation exchanging the two spinors, or equivalently to conjugation by an element of $O(2n)$ with
determinant minus one. Moreover, for $\fe_6$ one can choose a lift of the  Diagram automorphism by exchanging the
appropriate simple roots. One can check that the condition \eqref{eq:LNSV-Cond} is never
satisfied.

Finally we come to the special case of $\fd_4$. We view the root lattice as four-tuples
of integers with the sum of coordinates an even integer. Then in addition to the parity
involution $(x_1, x_2, x_3, x_4 ) \to (-x_1, x_2, x_3, x_4)$ there is a nontrivial
involution known as the Hadamard involution
\be
H = \half \begin{pmatrix}
1 & 1 & 1 & 1 \\
1 & -1 & 1 & -1 \\
1 & 1 & - 1 & -1 \\
1 & -1 & -1 & 1 \\
\end{pmatrix} \, .
\ee
There are root vectors such that $r \cdot H \cdot r$ is odd and hence the vector $p=(r;0)$
will satisfy \eqref{eq:LNSV-Cond}. Therefore, the orthogonal transformation $(p_L;p_R) \rightarrow (Hp_L; H p_R)$,
which is an involution of the Narain lattice will lift to an automorphism of the CFT which is either order four
or violates \eqref{eq:NaiveAct}.

\section{Cocycles At ADE Enhanced Symmetry Points}\label{sec:CocycleComments}

\subsection{Review Of Cocycles}\label{subsec:CocycleReview}

We first review the standard reason why cocycles are required in the construction of vertex operators for toroidal CFTs.
\footnote{See \cite{Goddard:1986bp}, section 6, for a particularly lucid account of the cocycles for chiral
vertex operator algebras associated with lattices. }
One might naively expect that under the state-operator correspondence the states $\vert p \rangle$ defining
a basis for $\IC[\Gamma]$ in \eqref{eq:CFT-statespace} correspond to the
vertex operator:
\be
V^{\rm naive}(p,z, \bar z)  :=  : e^{\I p X} : = : e^{\I (p_L X_L + p_R X_R) } :
\ee
where $p=(p_L; p_R)$ is the decomposition of $p$ into its left- and right-moving projections.
However the usual OPE
\be
V^{\rm naive}(p_1,z_1, \bar z_1) V^{\rm naive}(p_2,z_2, \bar z_2) = z_{12}^{p_L^1 \cdot p_L^2} \bar z_{12}^{p_R^1 \cdot p_R^2} : e^{\I (p_1 X(z_1, \bar z_1) + p_2 X(z_2, \bar z_2) ) } :
\ee
shows that the operators $V^{\rm naive}(p,z,\bar z)$ are not quite the right operators to use in a consistent CFT because they are not mutually local. In radial quantization we have the braiding relation:
\be\label{eq:BraidNaive}
V^{\rm naive}(p_1,z_1) V^{\rm naive}(p_2,z_2)  = e^{\I \pi p_1\cdot p_2}  V^{\rm naive}(p_2,z_2) V^{\rm naive}(p_1,z_1) \, .
\ee

The problem is with the factor  $e^{\I x_0 \cdot p} $ in the vertex operator. This is a shift operator
on $\IC[\Gamma]$ taking $L_{p'} \to L_{p+p'}$ and these operators generate   the commutative group
algebra $\IC[\Gamma]$.  In order to cancel the phase in \eqref{eq:BraidNaive} we introduce an extra operator $\hat c(p)$ on
$\IC[\Gamma]$ which is diagonal in the direct sum decomposition $\oplus_{p'} L_{p'}$ and acts as a multiplication
by a phase $\varepsilon(p,p')$ on $L_{p'}$ where the phases are valued in some subgroup $A\subset U(1)$. Then, if we define
\be
\hat C(p) := e^{\I x_0 \cdot p} \hat c(p)
\ee
these operators generate a noncommutative algebra
\be
\hat C(p_1) \hat C(p_2) = \varepsilon(p_1,p_2) \hat C(p_1+p_2) \, ,
\ee
where we have used the cocycle identity for $\varepsilon$.
The correct vertex operators:
\be
V(p,z,\bar z) :=  V^{\rm naive}(p,z,\bar z) \hat c(p)
\ee
will be mutually local if $\varepsilon$ satisfies the condition:
\be
s(p_1,p_2) := \frac{ \varepsilon(p_1,p_2)}{\varepsilon(p_2,p_1)} = e^{\I \pi p_1 \cdot p_2}
\ee
because
\be
\hat C(p_1) \hat C(p_2) = s(p_1,p_2) \hat C(p_2) \hat C(p_1) \, .
\ee

It is useful to interpret these formulae in terms of a central extension of the group $\Gamma$.
Associativity of the operators $\hat C(p)$ implies that $\varepsilon$ defines an $A$-valued
group cocycle on $\Gamma$, and hence defines a central extension:
\be\label{eq:Gamma-CE}
1 \rightarrow A \rightarrow \hat \Gamma \rightarrow \Gamma \rightarrow 1 \, .
\ee
This central extension acts on $\IC[\Gamma]$ with $A$ acting as scalars.
The central extension is characterized, up to isomorphism of central extensions,
by the commutator function $s(p^1, p^2)$. Changing $\varepsilon$ by a coboundary
corresponds to a redefinition of the the operators $\hat C(p)$ by a phase valued in
$A$, and the commutator function is gauge-invariant.
Note, however, that a choice of $A$ is part of the definition of the central
extension. Once $A$ has been chosen, valid coboundaries must be $A$-valued.
In much of the literature the group $A = \{ \pm 1 \} $ has been chosen, but
we will find that it is often more appropriate to let $A$ be the group of
fourth roots of unity.

\subsection{Detailed Cocycles For The $SU(2)$ Point}

We now demonstrate that the standard choice of cocycle is incompatible with $SU(2)$ symmetry at the
the $SU(2)$ enhanced symmetry point of a single Gaussian model. It is important to note that the inconsistency does
not arise at the level of vertex operators for the currents generating the affine $SU(2)$ algebra. It is well known that the
standard cocycle gives the correct commutation relations \cite{Frenkel:1980rn,Segal:1981ap,Goddard:1986bp}.
Rather as we will see,  the problem arises
in the OPE of currents with states transforming in the fundamental representation of $SU(2)$.

Our $SU(2)$ conventions are that we use anti-Hermitian generators
\be
T^a = - \frac{\I}{2} \sigma^a \qquad  [T^a, T^b] = \epsilon^{abc} T^c
\ee
so that the current $\times$ current OPE should be
\be
J^a(z_1) J^b(z_2) = \frac{ - \frac{k}{2} \delta^{ab}}{z_{12}^2} +
\frac{\epsilon^{abc}}{z_{12}} J^c(z_2) + \cdots \, .
\ee
where in general $k$ is the level and in our case $k=1$.

Now in the two-dimensional representation of $\fs\fu(2)$ we have
\be
T^+ := T^1 + \I T^2 = - \I \begin{pmatrix} 0 & 1 \\  0 & 0 \\ \end{pmatrix} , \qquad T^- := T^1 - \I T^2 = - \I \begin{pmatrix} 0 & 0 \\  1 & 0 \\ \end{pmatrix} \, .
\ee

So
\be
\begin{split}
[T^3, T^+ ] & = - \I T^+ \\
[T^3, T^- ] & = + \I T^- \\
[T^+, T^- ] & = - 2 \I T^3  \, .
\end{split}
\ee

Now $J^3 =  \p X_L /\sqrt{2}$ gives the OPE
\be
J^3(z_1) e^{\pm \I \sqrt{2} X_L} (z_2) = \frac{\mp \I}{z_{12}} e^{\pm \I \sqrt{2} X_L} (z_2)+ \cdots
\ee
so that up to normalization  $J^\pm(z) = : e^{\pm \I \sqrt{2} X_L} :\hat c(\pm (e+f))$ and
$\tilde J^\pm(z) = : e^{\pm \I \sqrt{2} X_R} :\hat c(\pm (e-f))$.

To determine the cocycle $\varepsilon(p_1,p_2)$ we first determine the usual constraint coming from the current-current OPEs
\be
J^+(z_1) J^-(z_2) = \varepsilon( e+f, -e-f)
\Biggl(  \frac{1}{z_{12}^2} +  \frac{ 2\I }{z_{12} } J^3(z_2) + \cdots  \Biggr) \, .
\ee
We thus require
\be
\varepsilon(e+f,-e-f) =   \varepsilon(-e-f, e+f) = \varepsilon(e-f,-e+f) =    \varepsilon(-e+f,e-f)  = - 1 \, .
\ee
%
%

However we should also demand that we  get the matrix elements of $T^\pm$ when
acting on the vertex operators with $n=\pm 1, w=0$ and $n=0, w= \pm 1$. These vertex operators
create states in the $(2,2)$ of $SU(2)_L \times SU(2)_R$.
Thus we consider
\be
V_{\epsilon_L, \epsilon_R}:=  : e^{\frac{\I}{\sqrt{2}} ( \epsilon_L X_L +
\epsilon_R X_R )}: \hat c_{\epsilon_L, \epsilon_R}
\ee
where different choices of signs $\epsilon_L, \epsilon_R$ give the four distinct vectors $\pm e, \pm f$.
We now compute the  OPE
\be
J^+(z_1) V_{-, \pm }(z_2)= -\frac{\I}{z_{12}} V_{+,\pm}(z_2) + \cdots
\ee
and so on.

Continuing in this way we find that \footnote{It is crucial in obtaining the signs below  to recall that the spectrum of the WZW model
consists of states with left-moving part in the representation ${\cal R}_\lambda$ corresponding to weights $\lambda -r$ with $r$ in the
root lattice and with right-moving part in the representation $\overline {{\cal R}_\lambda}$ with weights $-\lambda+r$. See equations \eqref{eq:FinDimRep} and \eqref{eq:FKS-basis}.}
\be \label{cocyceqns}
\begin{split}
\varepsilon( e+f, -f ) & = - \I  \qquad \varepsilon( e-f,  f )  =  \I \\
\varepsilon( e+f, -e ) & = - \I \qquad \varepsilon( e-f, -e )  =  \I \\
\varepsilon( -e-f, e ) & = - \I \qquad \varepsilon( -e+f, e )  =  \I \\
\varepsilon( -e-f, f ) & = - \I \qquad \varepsilon( -e+f, -f )  =  \I \\
\end{split}
\ee

To solve (\ref{cocyceqns}) we consider the general class of cocycles
\be
\varepsilon( n_1 e + w_1 f , n_2 e + w_2 f) =
e^{ \I \pi ( \alpha n_1 n_2 + \beta w_1 w_2+ \gamma n_1 w_2 + \delta w_1 n_2 ) }
\ee
with $\alpha, \beta, \gamma, \delta$ defined mod $2$
and impose (\ref{cocyceqns}) to obtain two solutions:
%
%
\begin{align}
\varepsilon_1(n_1 e + w_1 f , n_2 e + w_2 f) &= e^{ (\I \pi/2) ( -2w_1 w_2-  n_1 w_2 +  w_1 n_2 ) } \, , \\
\varepsilon_2(n_1 e + w_1 f , n_2 e + w_2 f) &= e^{ (\I \pi/2) ( -2 n_1 n_2+n_1 w_2 -  w_1 n_2 ) } \, .
\end{align}

These cocycles are in fact equivalent since they are related by a coboundary. Explicitly we have
\be
\varepsilon_2( n_1 e + w_1 f , n_2 e + w_2 f)= \varepsilon_1( n_1 e + w_1 f , n_2 e + w_2 f) e^{-i \pi(n_1-w_1)(n_2+w_2)}
\ee
and the factor on the right above is equal to
\be
e^{-i \pi(n_1+w_1)(n_2+w_2)}= e^{-2 \pi i p_L^1 \cdot p_L^2}= \frac{b(p^1+p^2)}{b(p_1)b(p_2)}
\ee
with $b(p)= {\rm exp}(-i \pi p_L^2)$.
From now on we work with the cocycle
$\varepsilon_1$.

We now show that this choice of cocycle ensures that the lift
of the Weyl group element is order four. In Appendix \ref{LLaut} we discuss a general formalism for lifting automorphisms
of abelian extensions of lattices.  In the notation used there we have a lattice extension $\hat \Gamma$ defined by the cocycle
$\varepsilon_1$ and can solve (\ref{xicon}) by choosing
\be\label{eq:LiftingFunction}
\xi_g(p) = {\rm exp} \left( (i \pi/2)(n+w)^2 \right) \, .
\ee
We then check that
\be
\xi_g(p) \xi_g(gp) = {\rm exp} \left( i \pi(n+w) \right)
\ee
which shows that the lift of the Weyl reflection is order four.

We close with two remarks. The first (pointed out to us by K. Wendland) is that
our cocycles do not satisfy the conndition $\varepsilon(-p,p)=1$ enforced in
\cite{Goddard:1986bp}. That condition is based on the choice of gauge $\varepsilon(0,p)=
\varepsilon(p,0)=1$ together with the condition $V(p)^\dagger = V(-p)$.
In fact, one could change the cocycle by a ($\IZ_4$-valued) coboundary to
enforce $\varepsilon(-p,p)=1$. Moreover, the Hermitian structure on the Hilbert space
of states and the state-operator correspondence is consistent with the more general
 Hermiticity condition to $V(p)^\dagger = (\varepsilon(-p,p))^{-1} V(-p)$,
 when $\varepsilon(-p,p)\not=1$.  The second remark is that the
 generalization of the above discussion to all the points $\Gamma(\fg)$ associated with simply
laced Lie algebras is not entirely trivial, and we hope to return to this question on a future
occasion.

%

\bigskip
\bigskip
\noindent
\textbf{\emph{Note added for v3}}: In the first two versions of this paper on the arXiv we
claimed that it is strictly necessary to modify the standard $\IZ_2$-valued cocycles in
vertex operator algebras to $\IZ_4$-valued cocycles in order to understand the nontrivial lifting
of $T$-duality discussed throughout the paper. This claim is erroneous. While the above formulae are correct,
so far as we know, one could perfectly well use the standard cocycle,
\be
\varepsilon(n_1 e + w_1 f, n_2 e + w_2 f) = (-1)^{n_1 w_2}
\ee
Indeed, this $\IZ_2$-valued cocycle can be obtained from $\varepsilon_1$ by a
$\IZ_4$-valued coboundary $b(ne + w f) = e^{\frac{\I \pi}{2} (w^2 - n w)}$ 
If we use the standard $\IZ_2$-valued cocycle
then we must  modify the lifting function \eqref{eq:LiftingFunction}
to
\be
\xi_g(n e + w f) = (-1)^{n (w+1)}
\ee
The matrix elements of $J^\pm$ acting on the half-spin modules are accounted for by a
simple rescaling of the vertex operators $V_{\epsilon_L, \epsilon_R}$ given above
(as is indeed implied by using the coboundary). We thank the referee for pointing this out.

\section{Criterion For Nontrivial Lifting}\label{sec:Involutions}

In this section we discuss the modular covariance approach to determining when
nontrivial elements  $g\in F(\Gamma)$ must lift to elements $\hat g \in \Aut(\CC_\Gamma)$
of twice the order of $g$, or  violate
\eqref{eq:NaiveAct}. We are discussing points in $\CN^{\rm ESP}$ that   typically
do not have non-abelian symmetry so we cannot use the crutch of the level one WZW model
for a non-abelian group. We will derive the criterion \eqref{eq:LNSV-Cond}.

\subsection{Inconsistency With Modular Covariance}\label{subsec:Involution-ModCov}

We begin by supposing that $g \in F(\Gamma)$ is an involution.
Suppose that there is a lift $\hat g$
so that $\langle \hat g \rangle \subset \Aut(\CC_{\Gamma})$ is isomorphic
to $\IZ_2$ and has an action \eqref{eq:NaiveAct} for $p\in \Gamma^g$. We are going to
show that if \eqref{eq:LNSV-Cond} is satisfied then there is an inconsistency
with modular covariance.

Using the methods of section \ref{sec:GenPF} below it is easy to see that
$Z(1,\hat g)$ has a $q$-expansion which has the form of a sum over $p\in (\Gamma^g)^\vee$ of
\be\label{eq:Coeff}
(q\bar q)^{-d/24}
\exp[ 2\pi \I (  n_-  \tau-  \tilde n_- \bar \tau)/16]  q^{\half p_L^2} \bar q^{\half p_R^2} \, ,
\ee
where $n_-$ is the number of twisted left-moving bosons and $\tilde n_-$ is the number of
twisted right-moving bosons,
times a power series in integral powers of $q^{1/2},\bar q^{1/2}$ with nonnegative integral
coefficients.  Under $\tau \to \tau + 2$
this transforms to
\be\label{eq:tau-add2}
(q\bar q)^{-d/24}
e^{\I \pi (n_- - \tilde n_-)/4}
\exp[ 2\pi \I (  n_-  \tau-  \tilde n_- \bar \tau)/16]  q^{\half p_L^2} \bar q^{\half p_R^2}
e^{\I \pi 2 p^2}
\ee
Now the key point is that (as we will show presently)
for every vector $p$ in the dual of the invariant lattice
  $2p^2 $ is an integer, but it can be even or odd.
If there are vectors for which it is odd,
 the sum over $p$ will produce a new function in the sense that:
\be
Z(1,\hat g)(\tau+2) \not = e^{\I \phi} Z(1, \hat g)(\tau)
\ee
for any phase $e^{\I \phi}$.
Therefore, if there are vectors in $(\Gamma^g)^\vee$ with $2p^2$ an odd integer
then in the $\hat g$-twisted sector $\hat g$ cannot be order two. One can then
check that modular covariance implies that $\hat g$ cannot be order two in the
untwisted sector either.

Note that for all  $p\in (\Gamma^g)^\vee$ it is true that
$4p^2$ is even. It follows that under a transformation $\tau \to \tau +4$ \eqref{eq:Coeff}
transforms to
\be\label{eq:tau-add4}
(q\bar q)^{-d/24}
e^{\I \pi (n_- - \tilde n_-)/2}
\exp[ 2\pi \I (  n_-  \tau-  \tilde n_- \bar \tau)/16]  q^{\half p_L^2} \bar q^{\half p_R^2}
\ee
and therefore this analysis of modular covariance indicates that it is consistent to assume
 that there is a lift $\hat g$ of $g$ that is order four.

Now we show that for $p \in (\Gamma^g)^\vee$,  $2p^2 $ is an integer, and in fact, the
existence of vectors such that it is an odd integer  is precisely equivalent to the condition \eqref{eq:LNSV-Cond}.
To prove this  let   $\fI = \Gamma^g$ be the sublattice of invariant
vectors.  Then, for every $v\in  \fI^\vee $ we have $v^2 \in \half \IZ$. Indeed we
have the usual decomposition of $\Gamma$ using glue vectors for
\be
(\Gamma^g) \oplus (\Gamma^g)^\perp := \fI \oplus \fN
\ee
and the discriminant groups of the invariant lattice $\fI$ and its orthogonal complement $\fN$
are isomorphic. So if $v \in \fI^\vee $ then there is a $u\in \fN^\vee$ with $w = v+ u \in \Gamma$.
Conversely, every $w\in \Gamma$ can be written in this form.
Therefore, since $g\cdot u = - u$ for $u \in \fN^\vee$,
\be
\begin{split}
w (1+g) w & = (v+u)^2 + (v+u)(v-u) \\
& = (v^2 + u^2) + (v^2 - u^2) \\
& = 2 v^2 \, . \\
\end{split}
\ee
Therefore, $2v^2 \in \IZ$, and, moreover, there is a $w\in \Gamma$ so that $w\cdot g \cdot w$ is odd
iff there is a vector $v\in \fI^\vee$ so that $2v^2$ is odd.

\subsection{Level Matching For Asymmetric Orbifolds By Involutions}

The discussion of the previous section is closely related to the level matching constraint in an asymmetric
orbifold using a nontrivial involution of the Narain lattice. Thus, consider
the asymmetric orbifold corresponding to the action   $X \to g X + s$ where
 $g^2=1$ and for simplicity we assume $g\cdot s = s$ so $2s \in \Gamma^g$.

The level matching constraint is
\be
2 \times \left(  \frac{n_-}{16} - \frac{\tilde n_-}{16} + \half (p+s)^2 \right) = 0~ \mod ~\IZ
\ee
where our convention for Narain lattices is  $p^2 = p_L^2- p_R^2$.
Here $p\in (\Gamma^g)^\vee$. Since $2s\in \Gamma^g$ this can be simplified to
\be\label{eq:LM-general}
  \frac{n_-}{8} - \frac{\tilde n_-}{8} + p^2 + s^2 = 0 ~\mod ~ \IZ \, .
\ee

When \eqref{eq:LM-general} is satisfied for \underline{every} vector $p\in (\Gamma^g)^\vee$
it follows (by subtracting the equation with $p=0$) that $p^2 = 0 ~\mod ~ \IZ$ for
every vector $p\in (\Gamma^g)^\vee$, hence $2p^2$ is always even and hence
for every vector  $P\in \Gamma$ we have  $P\cdot g \cdot P = 0 ~\mod ~2$. This is the
condition for the modular covariance of an order two lift $\hat g$ of $g$.
On the other hand, suppose we just know that \eqref{eq:LM-general} is satisfied for \underline{some} vector $p_0\in (\Gamma^g)^\vee$.
Then we can conclude, first of all that for \underline{every} vector $p\in (\Gamma^g)^\vee$ the
modular covariance condition for an order four lift $\hat g$  is satisfied:
\be\label{eq:LM-tau4}
  \frac{n_-}{4} - \frac{\tilde n_-}{4} + 2p^2 + 2s^2 = 0 ~\mod ~\IZ \, .
\ee
The reason is that we need only check that
\be
2p^2 - 2p_0^2 = 0 ~ \mod ~ \IZ
\ee
but we have seen that $2p^2 \in \IZ$ for every $p\in (\Gamma^g)^\vee$.

Now suppose that \eqref{eq:LM-general} is satisfied for \underline{some} vector $p_0\in (\Gamma^g)^\vee$ but not for
$p=0$. Then since $p_0^2 \in \half \IZ$ it must be that $2 p_0^2 $ is odd and hence there is some vector $w$ in $\Gamma$
satisfying \eqref{eq:LNSV-Cond}. As we have seen, this means there is no order two lift $\hat g$ consistent with modular
covariance. Moreover, even if $\hat g$ has order four, level matching would be violated by \underline{some} momentum sectors
in the first twisted sector (of the equivariant theory). Nevertheless, the relevant criterion for level matching
 for an order four element is that infinitely many states satisfy \eqref{eq:LM-general} for \underline{some}
 vector $p_0\in (\Gamma^g)^\vee$.

We conclude that the condition \eqref{eq:LNSV-Cond} should not be interpreted as a consistency condition for an orbifold
by a covering group of $\langle g\rangle$ based on considerations of level-matching in the first twisted sector.
We remark that the argument here did not
use any special properties of the formula for the right-moving ground state energy $\tilde n_-/16$ so exactly the
same reasoning will apply to the heterotic string.

\subsection{Twisted Characteristic Vectors}\label{subsec:TwistCharVect}

In preparation for section \ref{sec:GenPF} we note that the phase $e^{\I \pi  p \cdot g \cdot p}$
can be written as a character on the Narain lattice. That is, there is a vector $W_g \in \Gamma$
so that
\be\label{eq:TwistedCharVec}
e^{\I \pi  p \cdot g \cdot p} = e^{\I \pi p \cdot W_g} \, .
\ee
To prove this note  that, using that $g$ is an orthogonal involution:
\be
e^{\I \pi  (p_1+p_2) \cdot g (p_1 + p_2) }
=e^{\I \pi p_1 \cdot g  p_1 }
e^{\I \pi   p_2  \cdot g  p_2  }
\ee
so the map $\Gamma \to \IZ_2$ given by $p \mapsto e^{\I \pi  p \cdot g p } $
is a group homomorphism, and by Pontryagin duality
\footnote{The Pontryagin dual is $\Gamma^{PD} =\Hom(\Gamma, U(1))$
and here we are defining an order $2$ element of $\Gamma^{PD}$.
But for a locally compact abelian group $(G^{PD})^{PD} = G$ and
$G \times G^{PD} \to U(1)$ is a perfect pairing. In particular every homomorphism
in $\Gamma^{PD}$ is of the form $\chi_{\bar k} = p \mapsto e^{2\pi \I k \cdot p}$ where
$\bar k \in (\IR \otimes \Gamma^\vee)/\Gamma^\vee$ and $k$ is any lift
to $\IR \otimes \Gamma^\vee$. Since our homomorphism is two-torsion
and since $\Gamma$ is self-dual there is a vector $W_g \in \Gamma$ so
that $k = \half W_g$.  }
 there must be a vector $W_g\in \Gamma$ so that
\be
p \cdot g p =  p \cdot W_g ~ \mod ~ 2
\ee
thus proving equation \eqref{eq:TwistedCharVec}.
The vector $W_g$ is only defined up to addition by a vector in $2\Gamma$. It  is
the analog of an integral lift of a Stiefel-Whitney class.
\footnote{A characteristic vector
would satisfy $p^2 = p\cdot W ~\mod ~ 2$. It should not be confused with $W_g$.}

As an example, for the Gaussian model at the self-dual radius taking $g=\sigma_R$, which exchanges $e$ and $f$
 we have
\be
p \cdot g p = (ne + wf )\cdot (nf + w e) = n^2 + w^2 = n \pm w ~ \mod ~ 2 \, .
\ee
So this is indeed the same as the sign from before. We could take
\be
W_g =  \pm e \pm f
\ee
or any translate by an element of $2\Gamma$.
Note that we could choose a representative $W_ g=  e+f \in \Gamma^g$
which is orthogonal to $(\Gamma^g)^\perp$.

We can easily generalize this example by considering a product of Gaussian models,
all at the self-dual radius and with no $B$-field. Then $\Gamma$ is a direct sum of
$d$ copies of $\Gamma(r=1)$ with basis vectors $e_i, f_i$, $i=1,\dots, d$.
Then if $g: e_i \leftrightarrow f_i$  we have
\be
\Gamma^g  = \{   \sum_i  n_i  (e_i + f_i)  \}
\ee
\be
\Gamma^{g,\perp}  = \{   \sum_i  w_i  (e_i - f_i)  \}
\ee
and we can choose the representative
\be\label{eq:Wg-exple}
W_g = \sum_i (e_i + f_i)\in \Gamma^g
\ee

In section \ref{sec:GenPF} we will also need a
similar twisted characteristic vector relevant to the orbifold theory
by $\langle \hat g \rangle$.  We claim that there is a vector $W^{tw}_g\in \Gamma^g$ so that
\be
2p^2 = W^{tw}_g \cdot p ~ \mod ~ 2 \qquad \forall p \in (\Gamma^g)^\vee \, .
\ee
Moreover we claim there is a choice of
$W^{tw} _g\in \Gamma^g$ (as usual, modulo $2\Gamma^g$).
 In fact, we claim that there are
representatives of $W_g$ so that we can take $W_g^{tw}= W_g$.

To prove these statements about $W^{tw}_g$ we use the decomposition of
vectors $w\in \Gamma$ as $w = p + p'$ with $p \in (\Gamma^g)^\vee$
and $p' \in ((\Gamma^g)^\perp)^\vee$
and the identity
\be
w(1+g) w = 2 p^2
\ee
derived above. Now we note that
\be
(w_1 + w_2)(1+g)(w_1 + w_2) =
w_1 (1+g) w_1 + w_2 (1+g)w_2 +
[2 w_1 (1+g) w_2]
\ee
where we used the fact that $g$ is an involution. The term in
square brackets is even so
\be
(w_1 + w_2)(1+g)(w_1 + w_2) =
w_1 (1+g) w_1 + w_2 (1+g)w_2  ~\mod ~2
\ee
Then we see that $p \mapsto \exp[ \I \pi 2p^2 ] $ is
a group homomorphism  $(\Gamma^g)^\vee \rightarrow U(1)$
of order two so must be given by a homomorphism in the
torus $(\Gamma^g \otimes \IR)/\Gamma^g$ of order two.
In fact, we can do better: Since $W_g^{tw} \in \Gamma^g$,
if we write $w = p + p' \in (\Gamma^g)^\vee \oplus (\Gamma^{g,\perp})^\vee$
then (all equations taken modulo two):
\be
w\cdot W_g^{tw} = p\cdot W_g^{tw} = 2 p^2 = w (1+g) w= w\cdot W_g ~ \mod ~ 2
\ee
so we can take $W_g = W_g^{tw}$.

\subsection{Generalization To Elements Of Arbitrary Even Order}\label{subsec:ArbEvenOrder}

We can generalize the above discussion to elements $g\in F(\Gamma)$ of arbitrary order as
follows. We investigate modular covariance under the $\IZ_\ell$ subgroup generated by $g$.
In order to do this we need the generalization of equation \eqref{eq:Coeff} above.
The action of $g$ on $V_L\otimes \IC$ can be diagonalized so that it takes the form
\be
g \sim +1^{n_+} \oplus -1^{n_-} \oplus_{a} ( e^{2\pi \I \theta_a } \oplus e^{-2\pi\I \theta_a} )
\ee
where $a$ labels the eigenvalues of $g$ that are not $\pm 1$ so we can take $0 < \theta_a < 1$.
There is a similar diagonalization of  the action of $g$ on $V_R \otimes \IC$
with $n_+ \to \tilde n_+$, etc.  Then, assuming equation \eqref{eq:NaiveAct}
one can compute that $Z(1,\hat g)$ is a sum over terms $p\in (\Gamma^g)^\vee$:
\be
(q\bar q)^{-d/24} q^{E_0} \bar q^{\tilde E_0} q^{\half p_L^2} \bar q^{\half p_R^2} S(q^{1/\ell}, \bar q^{1/\ell} )
\ee
where $S$ is a series in nonnegative powers of $q^{1/\ell}$ and $\bar q^{1/\ell}$. The ground state energies are
\be
E_0 = \frac{n_-}{16} + \sum_a \half \theta_a (1- \theta_a)
\ee
\be
\tilde E_0 = \frac{\tilde n_-}{16} + \sum_{\tilde a} \half \tilde\theta_{\tilde a}  (1-   \tilde\theta_{\tilde a}  )
\ee
We now ask if it is consistent to assume that $\hat g$ has order $\ell$. Once again, the crucial point is that
for $p\in (\Gamma^g)^\vee$ we have
\be\label{eq:ell-p2}
\ell p^2 = P \cdot g^{\ell/2} \cdot P ~ \mod ~2
\ee
where $P$ is a vector $P\in \Gamma $ constructed below.  It follows that if $g$ has
even order $\ell$ and
\eqref{eq:LNSV-Cond} is satisfied, then modular covariance of $Z(1,\hat g)$ is violated \
for $\tau \to \tau + \ell$ if we apply \eqref{eq:NaiveAct}.  However, modular covariance is
consistent with the existence of a lift $\hat g$ of order $2\ell$, provided the
  $2\ell( E_0 - \tilde E_0 ) = 0~ \mod ~ 1$. On the other hand, if $ P \cdot g^{\ell/2} \cdot P = 0 ~ \mod ~2 $
  for all $P\in \Gamma$ then the existence of a lift $\hat g$ of $g$ of order $\ell$ is
  consistent with modular covariance, provided the standard level-matching constraint
  $\ell( E_0 - \tilde E_0 ) = 0~ \mod ~ 1$ is satisfied.

We now prove equation \eqref{eq:ell-p2}. We first note that for all $P\in \Gamma$, we
have
\be
P \cdot \left( 1+ g+ g^2 + \cdots  +  g^{\ell -1} \right) P =  \begin{cases}  0 ~ \mod~ 2 & \ell ~ {\rm odd} \\
P g^{\ell/2} P ~\mod ~2  & \ell ~ {\rm even} \\
\end{cases}
\ee
To prove this note that we can group terms so that
\be
\begin{split}
P \cdot \left( 1+ g+ g^2 + \cdots  +  g^{\ell-1} \right) P  & =     P^2 +  P\cdot ( g + g^{\ell-1}) P +
P \cdot (g^2 + g^{\ell-2})P + \cdots  \\
& +
\begin{cases}   P \cdot ( g^{(\ell-1)/2} + g^{(\ell+1)/2}) P & \ell ~ {\rm odd} \\
P g^{\ell/2} P & \ell ~ {\rm even} \\
\end{cases}
\\
\end{split}
\ee
Now $P^2$ is an even integer and
\be
 P g^k P +  P g^{\ell-k}P = P g^k P + P g^{-k} P = P g^k P + (g^k P) \cdot P = 2 P g^k P \in 2 \IZ \, .
 \ee
Therefore all the paired terms are even. The only thing left is the unpaired term when $\ell$ is
even.

Now, when we tensor over the complex numbers to consider $\Gamma$ embedded in the
complex vector space $\Gamma\otimes \IC$ we can apply projection operators
onto sublattices
transforming according to the irreducible characters of $\chi$ of $\IZ_\ell$:
\be
\otimes_{\chi\in Irrep(\IZ_\ell)} \fI_\chi
\ee
where $\fI_\chi = P_\chi \Gamma $ and $P_\chi$ is a projection operator.
 Then   every vector $P\in \Gamma$ has a decomposition
\be\label{eq:CompleteSum}
P = \sum_{\chi} p_\chi
\ee
with $p_\chi \in \fI_\chi$. Now note that
\be
\left( 1+ g+ g^2 + \cdots  +  g^{\ell-1} \right) p_\chi = \begin{cases}  \ell p_\chi  &  \chi =1 \\
0 & \chi \not= 1 \\
\end{cases}
\ee
Taking an inner produce with $P$ proves equation \eqref{eq:ell-p2} with $p_{\chi=1} = p$.
To complete the story we need to know that in fact every vector $p \in (\Gamma^g)^\vee$ has a completion \eqref{eq:CompleteSum}
with $P\in \Gamma$. To prove this we simply apply Nikulin's theorem to the primitively embedded
sublattice $\Gamma^g$.

These considerations suggest a natural conjecture for a canonical lift of $g$ to $\hat g$ in the automorphism group
of the CFT that acts as
\be\label{eq:GenRefConj-1}
\hat g \vert p \rangle  = e^{\I \pi \phi} \vert g\cdot p \rangle
\ee
where
\be\label{eq:GenRefConj-2}
\phi = \frac{1}{\ell} p \cdot \left( 1+ g+ g^2 + \cdots  +  g^{\ell-1} \right) p
\ee
We can check then that
\be
\hat g^\ell \vert p \rangle = \begin{cases}  \vert p \rangle & \ell ~ {\rm odd} \\
e^{\I \pi p g^{\ell/2} p } \vert p \rangle & \ell ~ {\rm even} \\
\end{cases}
\ee

As a check on this proposal consider the ADE point $\Gamma(\fg)$ and let
$g=(\sigma_{\alpha},1)$ be a left-moving reflection in a root. Acting on the
states of the form \eqref{eq:FKS-basis} our conjecture becomes:
\be\label{eq:CanLiftCheck}
\hat g \vert (\mu_L; \mu_R) \rangle = e^{-\frac{\I \pi}{2} (\alpha\cdot \mu_L)^2} \vert (\sigma_{\alpha}(\mu_L); \mu_R) \rangle
\ee
In particular, when $\sigma_{\alpha}(\mu_L) = \mu_L$ the eigenvalue is $+1$, exactly what we expect for the Tits lift.
Moreover, one can check explicitly for reflections in simple roots acting on the fundamental representation of $SU(N)$
that there is a basis of weight vectors
 such that equation \eqref{eq:CanLiftCheck} holds. Thus, our conjectured canonical lift appears to be
a generalization of the Tits lift for finite-dimensional groups acting on toroidal CFTs.

\section{General Discussion Of Partition Functions}\label{sec:GenPF}

In this section we consider a point in Narain moduli space with a nontrivial involution in $F(\Gamma)$
which satisfies the condition \eqref{eq:LNSV-Cond}. We assume that there is a lift of the involution
$\hat g$ so that
\be\label{eq:Assumption1}
\hat g \vert p \rangle = \vert p \rangle \qquad\qquad \forall ~ p \in \Gamma^g
\ee
\be\label{eq:Assumption2}
\hat g^2 \vert p \rangle = e^{\I \pi p \cdot W_g} \vert p \rangle  \qquad\qquad \forall p ~ \in \Gamma
\ee
where $p\cdot g \cdot p = p \cdot W_g ~\mod ~2$ for all $p \in \Gamma$ and we take $W_g \in \Gamma^g$
and not equivalent to zero.
We are interested in whether the orbifold by the $\IZ_4$ subgroup of $\Aut(\CC_{\Gamma})$ generated by $\hat g$ is consistent.
Just using the assumptions \eqref{eq:Assumption1} and \eqref{eq:Assumption2} and the method of modular orbits
 we will construct the partition function and in this section we will ask if the resulting partition function
 has a good $q$-expansion in the sense of section \ref{subsec:TechSumm}.
Of course, if we had an action of $\hat g$ on the full Hilbert space then it would
follow trivially that we have a good $q$ expansion, but we have not constructed a consistent vertex operator
algebra action on the various twisted sectors (including the untwisted sector)
and therefore it is useful to check whether the untwisted sector partition function
is consistent with an operator interpretation, which necessarily implies there is a good
$q$-expansion. In fact, we will find a new consistency condition, equation \eqref{eq:NewCC} below,
just from this necessary condition.

 To write the partition functions we will use the lattice theta functions defined in
 appendix \ref{app:ThetaFunctions}.
From \eqref{eq:Assumption1} we have:

\be\label{eq:Zg1}
Z(\hat g,1) = \frac{1}{\eta^{n_+}}\left( \frac{\vartheta_4(2\tau)}{\eta}\right)^{n_-}
 \frac{1}{\bar\eta^{\tilde n_+}} \left( \frac{\bar\vartheta_4(2\tau)}{\bar\eta}\right)^{\tilde n_-}
\Theta_{\Gamma^g}(\tau,0,0)
\ee
where $\Gamma^g$ is the sublattice of vectors fixed by $g$,
and $n_+ + n_- = d$. From \eqref{eq:Assumption2} we get:
\be
Z(\hat g^2 , 1) = \frac{1}{\eta^d} \frac{1}{\bar \eta^d}
\sum_{p \in \Gamma} q^{\half p_L^2} \bar q^{\half p_R^2} e^{2\pi \I ( p \cdot \half W_g)}
\ee

From \eqref{eq:Zg1} a modular transformation gives:
\be
Z(1, \hat g) =2^{-(n_- + \tilde n_-)/2} \vert \CD\vert^{-1/2}  \frac{1}{\eta^{n_+}}\left( \frac{\vartheta_2( \tau/2)}{\eta}\right)^{n_-}
 \frac{1}{\bar\eta^{\tilde n_+}} \left( \frac{\bar\vartheta_2( \tau/2)}{\bar\eta}\right)^{\tilde n_-}
\Theta_{(\Gamma^g)^\vee}(\tau; 0, 0)
\ee
where $\CD$ is the discriminant group of $(\Gamma^g)^\vee$. Now we want to average this over shifts of $\tau$
to construct the partition function in the first twisted sector.

When checking that we get good $q$-expansions it will be useful to define
\be
\begin{split}
\vartheta_2(\tau/2) & = q^{1/16} \sum_{n\in\IZ} e^{\I \pi \frac{\tau}{2} (n^2 + n)}
= 2 q^{1/16} \left( 1+ \sum_{n=1}^\infty  q^{\frac{n(n+1)}{4}} \right)\\
& := 2 q^{1/16} S(\tau)\\
\end{split}
\ee
Note that $S$ is a power series in positive powers of $q^{1/2}$ with positive integral coefficients.
In these terms we can write:
\be
Z(1, \hat g) =
D (B_+ \bar B_+)^{d} q^{n_-/16} \bar q^{\tilde n_-/16}  S^{n_-} \bar S^{\tilde n_-}
\Theta_{(\Gamma^g)^\vee}(\tau; 0, 0)
\ee
where $B_+ = 1/\eta$ and
\be
D := \sqrt{ \frac{ 2^{(n_- + \tilde n_-)/2} }{\vert \CD \vert} }
\ee
is an integer, according to \cite{LepowskyCalculus, Narain:1986qm}.
%
%

Next,   $\tau \to \tau + 2$ gives the partition function:
\be
Z(\hat g^2, \hat g) =e^{\I \pi (n_- - \tilde n_-)/4}
2^{-(n_- + \tilde n_-)/2} \vert \CD\vert^{-1/2}  \frac{1}{\eta^{n_+}}\left( \frac{\vartheta_2( \tau/2)}{\eta}\right)^{n_-}
 \frac{1}{\bar\eta^{\tilde n_+}} \left( \frac{\bar\vartheta_2( \tau/2)}{\bar\eta}\right)^{\tilde n_-}
\Theta_{(\Gamma^g)^\vee}(\tau;\alpha,0)
\ee
where $\alpha = - \half W_g$.  Now we can again use a modular transform to get
\be\label{eq:Zggsq}
Z(\hat g, \hat g^2) =e^{\I \pi (n_- - \tilde n_-)/4}
   \frac{1}{\eta^{n_+}}\left( \frac{\vartheta_4( 2\tau )}{\eta}\right)^{n_-}
 \frac{1}{\bar\eta^{\tilde n_+}} \left( \frac{\bar\vartheta_4( 2\tau )}{\bar\eta}\right)^{\tilde n_-}
\Theta_{\Gamma^g }(\tau; 0 ; \half W_g )
\ee
Modular invariance (and level matching) requires $n_-  -  \tilde n_- = 0 ~ \mod ~4$.
Equation \eqref{eq:Zggsq} shows that if
$n_-  -  \tilde n_- = 4 ~ \mod ~ 8$ then we get bad signs that can potentially
spoil the operator interpretation. Level matching is not strong enough to guarantee
a good $q$-expansion.

To compute the partition function in the $\hat g^2$-twisted sector we begin with
\be
Z(\hat g^2 , 1) = \frac{1}{\eta^d} \frac{1}{\bar \eta^d}
\sum_{p \in \Gamma} q^{\half p_L^2} \bar q^{\half p_R^2} e^{2\pi \I ( p \cdot \half W_g)}
\ee
and then
\be
Z(1, \hat g^2) =  \frac{1}{\eta^d} \frac{1}{\bar \eta^d}
\sum_{p \in \Gamma} e^{\I \pi \tau ( p_L + \half W_{g,L})^2}  e^{-\I \pi \bar\tau
(p_R + \half W_{g,R})^2 }
\ee
Now taking $\tau \to \tau +1$ we get:
\be
Z(\hat g^2, \hat g^2) =  e^{2\pi\I \frac{W_g^2}{8}} \frac{1}{\eta^d} \frac{1}{\bar \eta^d}
\sum_{p \in \Gamma} e^{\I \pi \tau ( p_L + \half W_{g,L})^2}  e^{-\I \pi \bar\tau
(p_R + \half W_{g,R})^2 }  e^{\I \pi p \cdot W_g}
\ee

We now have all the ingredients to write the full partition functions.
We would like to check that   all coefficients  in the   $q$, $\bar q$ -expansion in
all four sectors  are nonnegative integers.

We first consider the untwisted sector and this is just:
\be\label{eq:Untwist}
\begin{split}
Z(\CH_1) & = \frac{1}{4} \frac{1}{\eta^d \bar \eta^d}
\Biggl[ \sum_{p\in \Gamma} e^{\I \pi \tau p_L^2 - \I \pi \bar \tau p_R^2 }
\left( 1+ e^{\I \pi p \cdot W_g } \right) \\
& + 2 (\vartheta_4(2\tau))^{n_-}  (\bar\vartheta_4(2\tau))^{\tilde n_-}
\sum_{p\in \Gamma^g} e^{\I \pi \tau p_L^2 - \I \pi \bar \tau p_R^2 }
  \Biggr] \\
\end{split}
\ee
The potential problem here are the minus signs from the factors $\vartheta_4$
and $\bar \vartheta_4$. Also the coefficients are potentially half-integral.
(The vacuum is easily seen to have degeneracy $1$.)

We claim there is a good operator interpretation. To show this define
$\Gamma_0 :=  \Gamma^g \oplus \Gamma^{g,\perp}$.
Then we can write
\be
\Gamma = \amalg_{i=0}^{d}   (\Gamma_0 + \gamma_i)
\ee
where the glue vectors $\gamma_i$ project to representatives of the discriminant
group. Then we can write
\footnote{Note that this step uses the fact that if $p_1\in \Gamma^g$ and $p_2 \in (\Gamma^g)^\vee$
then not only is $p_1 \cdot p_2 =0$, but also  $p_{1,L} \cdot p_{2,L} = 0$. This follows since
$g(p_L;p_R) = (g_L p_L; g_R p_R)$  with $g_L, g_R$ both involutions. We thank K. Wendland for
a clarifying remark on this point.}
\be\label{eq:Untwist2}
\begin{split}
Z(\CH_1) & = \frac{1}{2} \frac{1}{\eta^d \bar \eta^d}
\Biggl[ ( \sum_{p\in \Gamma^{g,\perp}} e^{\I \pi \tau p_L^2 - \I \pi \bar \tau p_R^2 }
+  (\vartheta_4(2\tau))^{n_-}  (\bar\vartheta_4(2\tau))^{\tilde n_-}  )
\sum_{p\in \Gamma^g} e^{\I \pi \tau p_L^2 - \I \pi \bar \tau p_R^2 }   \\
& +  \sum_{\gamma_i\not=0} \frac{ 1+ e^{\I \pi \gamma_i \cdot W_g} }{2}
\sum_{p\in \Gamma_0 + \gamma_i } e^{\I \pi \tau p_L^2 - \I \pi \bar \tau p_R^2 }
  \Biggr] \\
\end{split}
\ee
Regarding the sum over glue vectors we note that if
$\half ( 1+ e^{\I \pi \gamma_i \cdot W_g}) =+1$ then
$\half ( 1+ e^{-\I \pi \gamma_i \cdot W_g}) =+1$ so we can
pair the terms with $p$ and $-p$ and that cancels the overall
factor of $1/2$ and yields a series with nonnegative integer coefficients.
If $-\gamma_i = \gamma_i ~ \mod ~ \Gamma_0 $ then there is only one term in the
sum over $\gamma_i$ but then $\sum_{p\in \Gamma_0 + \gamma_i } e^{\I \pi \tau p_L^2 - \I \pi \bar \tau p_R^2 }$
has degeneracies which are multiples of $2$.

For the remaining terms it would  suffice to prove that
\be\label{eq:TransTheory}
\half \frac{1}{\eta^{n_-} \bar\eta^{\tilde n_-}}
\Biggl(   \sum_{p\in \Gamma^{g,\perp}} e^{\I \pi \tau p_L^2 - \I \pi \bar \tau p_R^2 }
+  (\vartheta_4(2\tau))^{n_-}  (\bar\vartheta_4(2\tau))^{\tilde n_-}   \Biggr)
\ee
is a positive $q, \bar q$ expansion with nonnegative integer coefficients.
 But note that  the
lattice $\Gamma^{g,\perp}$ is even and signature $(n_-; \tilde n_-)$.
This expression is manifestly the untwisted sector partition function
of a system of bosons on $\Gamma^{g,\perp}$ with the orbifold action
$p \to -p $. It therefore has an operator interpretation.

The partition function in the $\hat g^2$-twisted sector is
\be\label{eq:g2Twist}
\begin{split}
Z(\CH_{\hat g^2}) & = \frac{1}{4} \frac{1}{\eta^d \bar \eta^d}
\Biggl[ \sum_{p\in \Gamma} e^{\I \pi \tau (p+\half W_g)_L^2 - \I \pi \bar \tau
(p+\half W_g)_R^2 }
\left( 1+ e^{2\pi \I \frac{W_g^2}{8}} e^{\I \pi p \cdot W_g } \right) \\
& + e^{\I \pi \frac{(n_-  - \tilde n_-)}{4} }  (\vartheta_4(2\tau))^{n_-}  (\bar\vartheta_4(2\tau))^{\tilde n_-}
\sum_{p\in \Gamma^g} e^{\I \pi \tau (p+\half W_g)_L^2 - \I \pi \bar \tau (p+\half W_g)_R^2 }
\left( 1+ e^{2\pi \I \frac{W_g^2}{8}} e^{\I \pi p \cdot W_g } \right) \Biggr] \\
\end{split}
\ee
Now again we have  to worry about   potential signs and half-integers.

Now to make progress note that
 $e^{2\pi \I \frac{W_g^2}{8}} $  is always
 a fourth root of unity since $W_g \in \Gamma^g$ is in an even lattice.
 We will now argue that
 $e^{2\pi \I W_g^2/8}$ should be a sign.
 Let us define
 \be
 \xi:= e^{2\pi \I W_g^2/8}\qquad\qquad \xi' = e^{\I \pi (n_- - \tilde n_-)/4}
 \ee
We know that $\xi'$ is $\pm 1$ by the cancellation of modular anomalies.

In analogy to \eqref{eq:Untwist2}, we can write \eqref{eq:g2Twist}
 as
\be\label{eq:g2Twist2}
\begin{split}
Z(\CH_{\hat g^2} ) & = \half \frac{1}{\eta^d \bar \eta^d}
\Biggl[ \frac{1+\xi}{2} \biggl\{ \sum_{p\in \Gamma^{g,\perp}} e^{\I \pi \tau p_L^2 - \I \pi \bar \tau p_R^2 }
+ \xi'  (\vartheta_4(2\tau))^{n_-}  (\bar\vartheta_4(2\tau))^{\tilde n_-}  \biggr\}
\left(
\sum_{p\in \Gamma^g+\half W_g } e^{\I \pi \tau p_L^2 - \I \pi \bar \tau p_R^2 } \right)  \\
& +  \sum_{\gamma_i\not=0} \frac{ 1+\xi  e^{\I \pi \gamma_i \cdot W_g} }{2}
\sum_{p\in \Gamma_0 + \gamma_i +\half W_g } e^{\I \pi \tau p_L^2 - \I \pi \bar \tau p_R^2 }
  \Biggr] \\
\end{split}
\ee

If $\xi$ is $\pm \I$ then it is clear that we will not get an integral expansion
in \eqref{eq:g2Twist2}. For example, we could choose $W_g$ to be minimal length among its representatives
and then the leading term in the $q$ expansion will involve
\be
\half (1+ \xi)(1+\xi') e^{\I \pi \tau W_{g,L}^2/4} e^{-\I \pi \bar\tau W_{g,R}^2/4}
\ee
If $\xi'=1$ then it is clear that we cannot have $\xi = \pm \I$. If $\xi'=-1$ we must look at the
next-to-leading terms and again it is clear we cannot have $\xi = \pm \I$.
Therefore we must have $\xi^2 =1$. Thus a consistency condition for asymmetric orbifolds is the requirement
that
\be\label{eq:NewCC}
\xi^2 = e^{2 \pi \I W_g^2/4} = 1 \, .
\ee
We believe this condition has not appeared in the literature before.

Given that $\xi^2 = 1$ the argument that $Z(\CH_{\hat g^2})$ has a good $q$-expansion
is very similar to that for the untwisted sector. In the
sum over $\gamma_i$ we pair up terms with $\gamma_i$ and $-\gamma_i - W_g$ (and when these are
the same in the discriminant group then the shifted theta function has even degeneracies).
What we need to check is that
\be\label{eq:TransTheory2}
\half \frac{1}{\eta^{n_-} \bar\eta^{\tilde n_-}}
\Biggl(   \sum_{p\in \Gamma^{g,\perp}} e^{\I \pi \tau p_L^2 - \I \pi \bar \tau p_R^2 }
+ \xi' (\vartheta_4(2\tau))^{n_-}  (\bar\vartheta_4(2\tau))^{\tilde n_-}   \Biggr)
\ee
is a positive $q, \bar q$ expansion with nonnegative integer coefficients.
Again as with \eqref{eq:TransTheory}
we interpret this in terms of a system of bosons on $\Gamma^{g,\perp}$ with the orbifold action
$p \to -p $. It therefore has an operator interpretation. Depending on $\xi'$ we might be
  projecting to the anti-invariant
subspace, but it still has a good $q$-expansion.

Finally, we must check that the operator interpretation is sensible in
the $\hat g$-twisted sector $\CH_{\hat g}$:
\be
Z(\CH_{\hat g}) = \frac{1}{4} \biggl(  Z(1, \hat g)(\tau) + Z(1, \hat g)(\tau+1) +
Z(1, \hat g)(\tau+2)  + Z(1, \hat g)(\tau+3)  \biggr)
\ee
To do this we return to the equation:
\be
Z(1, \hat g) =
D (B_+ \bar B_+)^{d} q^{n_-/16} \bar q^{\tilde n_-/16} S^{n_-} \bar S^{\tilde n_-}
\Theta_{(\Gamma^g)^\vee}(\tau; 0, 0) \, .
\ee
Now write the terms in the theta function as a sum over
\be
q^{\half p^2} (q \bar q)^{\half p_R^2} \, .
\ee
But $q^{\half p^2}$ is $q^\mu$ where $\mu \in \frac{1}{4} \IZ$. Similarly we can write:
\be
q^{n_-/16} \bar q^{\tilde n_-/16} = q^{(n_- - \tilde n_-)/16} (q \bar q)^{\tilde n_-/16} \, .
\ee
Since $q\bar q$ is inert under $\tau \to \tau + 1$, when $n_- - \tilde n_- =0 ~ \mod ~4$
 we can write the whole partition function in the form:
\be
Z(1, \hat g)(\tau) = D \sum_{\mu,\nu\in \frac{1}{4} \IZ} \wp_{\mu,\nu}( (q\bar q)^{1/\ell}) q^{\mu} \bar q^{\nu}
\ee
where $\ell$ is some integer (for rational theories) and $ \wp_{\mu,\nu}(x)$ is a power series in $x$ with
positive integer coefficients.  Now the sum over shifts of $\tau$ just projects to the subset of terms
with $\mu - \nu = 0 ~ \mod ~ 1$. This concludes the proof. $\spadesuit$

To conclude we remark that the consistency condition \eqref{eq:NewCC}
is satisfied for \eqref{eq:Wg-exple} since   $W_g^2/4 = (\tilde n_- - n_-)/4$
in this example.

\section*{Acknowledgements}

We would like to thank T. Banks, D. Freed, D. Friedan, M. Gaberdiel, D. Gaiotto, P. Goddard, Y. Z. Huang, J. Lepowsky,
K.S. Narain, I. Runkel, C. Schweigert, G. Segal, A. Taormina, W. Taylor,
 A. Tripathy, C. Vafa,  and R. Volpato for helpful discussions and correspondence.  We are very grateful to J. Cushing and K. Wendland for providing detailed feedback that helped us correct minor errors and improve the presentation. We
are especially grateful to N. Seiberg for some collaboration on these matters and for the essential remark
that $T$ duality of the Gaussian model is the lifting of a $180$ degree rotation.
This collaboration owes a great deal to the hospitality of the Aspen Center for Physics  (under NSF Grant No. PHY-1066293).
JH acknowledges support from the NSF \footnote{Any opinions, findings, and conclusions or recommendations expressed in this material are those of the author(s) and do not necessarily reflect the views of the National Science Foundation.} under grant PHY 1520748 and from the Simons Foundation (\#399639).  GM thanks   Institute for Advanced Study in Princeton for support from   the IBM Einstein Fellowship of the Institute for Advanced Study.
GM also acknowledges support by the DOE under grant
DOE-SC0010008 to Rutgers University.

\appendix

\section{Automorphism Groups Of Extensions Of Lattices} \label{LLaut}

As is well known, and as discussed in section \ref{sec:CocycleComments},  locality of the OPE for vertex operators requires that we consider a group extension of the momentum lattice $\Gamma$.
This takes the form
\be
1 \rightarrow A \rightarrow \widehat \Gamma \rightarrow \Gamma \rightarrow 0 \, .
\ee
As discussed in the text, much of the literature takes $A$ to be isomorphic to $\IZ/2 \IZ$, but we have argued that
group invariance at enhanced symmetry points requires $A=\IZ/4 \IZ$. Here we just assume that $A$ is a finite abelian group.
We write the group law in $\Gamma$ additively and the group law in $A$ multiplicatively. We now discuss how to lift
automorphisms of $\Gamma$ to automorphisms of $\widehat{\Gamma}$.

We begin by constructing a group $\widehat{\Aut(\Gamma)}$ that is a subgroup of the group of automorphisms of $\hat \Gamma$
and covers the action of $\Aut(\Gamma)$ on $\Gamma$. Our group will fit in an extension of the form
\be\label{eq:HomSeq}
1 \rightarrow {\rm Hom}(\Gamma,A) \rightarrow \widehat{\Aut(\Gamma)} \rightarrow \Aut(\Gamma) \rightarrow 1  \, .
\ee
We denote elements of  $\Gamma$ by $p$, elements of $A$ by $a$ and the
action of $g \in \Aut(\Gamma)$ on $p \in \Gamma$ by $ g p$.
Elements of $\hat \Gamma$ are pairs $(a,p)$ with composition law
\be
(a_1,p_1) \cdot (a_2,p_2) = (a_1 a_2 \varepsilon(p_1,p_2), p_1+p_2)
\ee
with $\varepsilon$ a cocycle.  For each $g \in \Aut(\Gamma)$ we wish to define
an element $T_g \in {\rm Aut}(\hat \Gamma)$ of the form
\be
T_g(a,p)= (a \xi_g(p), g p) \, ,
\ee
where $\xi_g$ is a function from $\Gamma$ to $A$.
Demanding that $T_g \in \Aut(\hat \Gamma)$ gives a constraint on $\xi_g$:
\be \label{xicon}
\frac{\xi_g(p_1+p_2)}{\xi_g(p_1) \xi_g(p_2)}=\frac{\varepsilon(g p_1, g p_2)}{\varepsilon(p_1,p_2)} \, .
\ee
For each $g\in F(\Gamma)$ we choose a solution of \eqref{xicon} (we assume it exists).  Note that given one solution
we can multiply $\xi_g$ by any element $\ell_g \in \Hom(\Gamma,A)$ to produce another solution. Note that if
we change $\varepsilon$ by a coboundary $b$ then $\xi_g$ will be replaced by
\be\label{eq:CoboundaryTmn}
\tilde \xi_g(p) = \xi_g(p) \frac{b(p)}{b(g p) }
\ee

The set of operators $T_g$ for $g \in \Aut(\Gamma)$ generate a subgroup of $\Aut(\hat \Gamma)$
which is an extension of $\Aut(\Gamma)$. Now a small computation shows that
%
%
%
%
%
\be
T^{-1}_{g_1 g_2} \circ T_{g_1} \circ T_{g_2}(a,p) = \left( a \cdot (\xi_{g_1 g_2}(p))^{-1}\xi_{g_2}(p) \xi_{g_1}(g_2 p),p \right) \, .
\ee
Now for each $g_1, g_2$ define a function $\ell_{g_1,g_2}: \Gamma \to A$ by
\be
\ell_{g_1,g_2}(p):= (\xi_{g_1 g_2}( p))^{-1} \xi_{g_2}(p) \xi_{g_1}(g_2 p) \, .
\ee
Another short computation using (\ref{xicon}) shows  that $\ell_{g_1,g_2}$ is a linear function:
\be
\ell_{g_1,g_2}(p_1+p_2) = \ell_{g_1,g_2}(p_1) \ell_{g_1,g_2}(p_2)
\ee
and hence $\ell_{g_1,g_2} \in \Hom(\Gamma, A)$. Now, for each $\ell \in \Hom(\Gamma, A)$ define
an automorphism $L_{\ell} \in \Aut(\hat \Gamma)$ by
\be
L_\ell (a, p) = (a \ell(p), p)
\ee
Applying \eqref{xicon} shows that $L_\ell$ is indeed an automorphism of $\hat \Gamma$.
We have shown that
\be\label{eq:TT}
T_{g_1} \circ T_{g_2} = T_{g_1g_2}\circ L_{\ell_{g_1,g_2}}
\ee
But now note that $\Hom(\Gamma, A)$ is itself a group under pointwise multiplication:
$(\ell_1\cdot \ell_2)(p):= \ell_1(p) \ell_2(p)$ where the RHS is defined by multiplication
in $A$ and clearly
\be\label{eq:LL}
\begin{split}
L_{\ell_1} \circ L_{\ell_2} & = L_{\ell_1 \cdot \ell_2} \\
\end{split}
\ee
Moreover, $\Aut(\Gamma)$ acts on $\Hom(\Gamma,A)$ via   $g \cdot \ell(p):= \ell(g\cdot p)$
and one can check that
\be\label{eq:LT-TL}
\begin{split}
L_{\ell} \circ T_g & = T_g \circ L_{g\cdot \ell}  \,  . \\
\end{split}
\ee
The equations \eqref{eq:TT}, \eqref{eq:LL}, and \eqref{eq:LT-TL} show that the
set of automorphisms
\be
\widehat{\Aut(\Gamma)}:= \{ T_{g,\ell} := T_g L_\ell \}
\ee
labeled by $(g,\ell) \in \Aut(\Gamma) \times \Hom(\Gamma,A)$   form a group with multiplication law:
\be
T_{g_1,\ell_1} T_{g_2,\ell_2} = T_{g_1 g_2, \ell_{g_1,g_2} \cdot (g_2\cdot \ell_1) \cdot \ell_2}
\ee
The injection $\ell \mapsto L_\ell$ and projection $T_{g} L_\ell \mapsto g$ show that the group fits in the
exact sequence \eqref{eq:HomSeq}.

Now we can restrict to the subgroup of $\widehat{\Aut(\Gamma)}$ that projects to
$F(\Gamma) = \Aut(\Gamma) \cap \left( O(d)_L \times O(d)_R \right)$. Or we can even
restrict to a subgroup of $F(\Gamma)$. The main example in the text is the case
where $g \in F(\Gamma)$ is a nontrivial involution that generates a $\IZ_2$-subgroup of
$F(\Gamma)$. In this case the square of $T_g$ is given by
\be
T_g \cdot T_g (a,p) = (a \xi_g(p) \xi_g(g p),p)
\ee
The element $\xi_g(p) \xi_g(gp)$ is invariant under a change of cocycle
by a coboundary, as one easily checks using \eqref{eq:CoboundaryTmn}.
In this sense it is gauge invariant.  Thus, $T_g$ squares to the identity only if
\be
\xi_g(p) \xi_g(gp)=1 \, .
\ee
In the examples of section \ref{sec:CocycleComments}  we find rather that $\xi_g(p)\xi_g(gp)$ is a $\IZ_2$-valued linear
function that is, moreover, $g$-invariant, so in this case the restriction
of the extension to $\langle g \rangle \subset F(\Gamma)$ is just an extension of $\IZ_2$ by $\IZ_2$,
consistent with the $\IZ_4$ lift we found using $SU(2)$ invariance.


\section{Transformation Of Boundary Conditions} \label{bctrans}

Suppose our field on the torus has twisted boundary conditions
\be
\begin{split}
X(\sigma_1 +1, \sigma_2) & = g_s \cdot X(\sigma_1, \sigma_2) \\
X(\sigma_1 , \sigma_2+1) & = g_t \cdot X(\sigma_1, \sigma_2) \\
\end{split}
\ee
with modular parameter:
\be
\vert dz\vert^2 = \vert d\sigma_1 + \tau d \sigma_2 \vert^2
\ee
For an $SL(2,\IZ)$ transformation define
\be
\begin{split}
\sigma_1 & = d \sigma_1' + b \sigma_2' \\
\sigma_2 & = c \sigma_1' + a \sigma_2' \\
\end{split}
\ee
so that
\be
\tau' = \frac{a \tau + b }{ c \tau +d }
\ee

Now, under $(\Delta \sigma_1' = 1 , \Delta \sigma_2'=0)$
we have $(\Delta\sigma_1 = d , \Delta \sigma_2   = c )$ etc. So
\be
X(\sigma_1' + 1, \sigma_2') = g_s^d g_t^c X(\sigma_1', \sigma_2')
\ee
and so on.  In this way we derive
\be
Z(g_s^b g_t^a, g_s^d g_t^c; \tau') = Z(g_t, g_s ; \tau)
\ee
(This just says we should get the same answer working in
$\sigma'$-variables.)
Making a few trivial change of variables this means:
\be\label{eq:BC-TMN}
Z(g_t, g_s; \tau') = Z(g_s^{-b} g_t^{d}, g_s^{a} g_t^{-c}; \tau)
\ee

Note that the action on functions of $\tau$ must descend to $PSL(2,\IZ)$, but
  the action on the boundary conditions:
  \be
 (g_t, g_s) \rightarrow (g_s^{-b} g_t^{d}, g_s^d g_t^{-c})
 \ee
 does not descend. Therefore equation \eqref{eq:BC-TMN} only makes sense if:
 \be
Z(  g_t^{-1}, g_s^{-1} ; \tau) = Z(  g_t , g_s  ; \tau)
 \ee
 for all commuting pairs $g_s, g_t$.

\section{Theta Functions}\label{app:ThetaFunctions}

Suppose that $\IR^{b_+, b_-}$ is Euclidean space with quadratic form $\eta_{AB} = (+1^{b_+}, -1^{b_-})$.
We use indices $a,b,\dots = 1, \dots , b_+$ for the Euclidean coordinates on the positive definite space
and $s,t,\dots = 1, \dots, b_-$ for Euclidean coordinates on the negative definite space, while $A,B,\dots $
run from $1$ to $d:=b_+ + b_-$.

Now suppose that $\Lambda \subset \IR^{b_+,b_-}$ is an embedded lattice. It is the integral span of
vectors $e^{A}_{~~i}$ so we have vectors with coordinates $x^A$:
\be
x^A = \sum_{i=1}^d  n^i e^{A}_{~~i}  \qquad A=1,\dots, d
\ee
The Gram matrix is
\be
G_{ij} = e^{A}_{~~i} \eta_{AB} e^{B}_{~~j}
\ee
At this point we are not making any integrality assumptions about $G_{ij}$. It is just a nondegenerate symmetric
real matrix. We consider the theta function:
\be
\begin{split}
\Theta_{\Lambda}(\tau, \alpha, \beta) &:=\sum_{\lambda\in \Lambda} e^{\I \pi \tau (\lambda+\beta)_+^2 + \I \pi \bar\tau (\lambda+\beta)_-^2
- 2\pi \I (\lambda + \half \beta, \alpha)} \\
& = \sum_{n^i\in \IZ} e^{ (n^i+\beta^i)( n^j+\beta^j) Q_{ij}(\tau) - 2\pi \I (n^i + \half \beta^i) \alpha^j G_{ij} }  \\
\end{split}
\ee
with
\be
Q_{ij}(\tau) = \sum_{a=1}^{b_+} \I \pi \tau  e^{a}_{~i}  e^{a}_{~j} -\sum_{s=1}^{b_-} \I \pi \bar\tau  e^{s}_{~i}  e^{s}_{~j}
\ee
The Poisson summation formula gives:
\be\label{eq:Mod-Tmn}
\Theta_{\Lambda}(-1/\tau, \alpha, \beta) = (-\I \tau)^{b_+/2} (\I \bar \tau)^{b_-/2} \vert \det e^{i}_{~A} \vert
\Theta_{\Lambda^\vee}(\tau, \beta, - \alpha)
\ee
where $\Lambda^\vee$ is the lattice spanned by the vectors with coordinates
\be
x_A =\sum_{i=1}^d  m_i e^{i}_{~A}
\ee
with $m_i \in \IZ$ and $e^{i}_{~A}$  is the inverse matrix of
$e^{A}_{~i}$ . Note that consistency with making two S transformations requires
\be
\Theta_{\Lambda}(\tau, \alpha, \beta)= \Theta_{\Lambda}(\tau, - \alpha, - \beta)
\ee
which is indeed the case.

Up to this point we have \underline{not} assumed $G_{ij}$ is an integral matrix. In particular $\Theta_{\Lambda}(\tau, \alpha, \beta)$
does not have any special properties under $\tau \to \tau + 1$.  Now assume that $G_{ij}$ is an integral matrix. Then
\be
\vert \det e^{i}_{~A} \vert = \sqrt{\vert \det G^{ij} \vert } = \frac{1}{\sqrt{\vert \det G_{ij} \vert}} = \frac{1}{\sqrt{\vert \CD \vert} }
\ee
and $\vert \CD \vert$ is the order of the discriminant group. So for integral lattices we have the S-transformation
\be
\Theta_{\Lambda}(-1/\tau,\alpha,\beta):= (-\I \tau)^{b_+/2} (\I \bar \tau)^{b_-/2} \vert \CD \vert^{-1/2}
\Theta_{\Lambda^\vee}(\tau,\beta,-\alpha)
\ee

In the text we sometimes use the standard theta functions:
\be
\begin{split}
\vartheta_2 & = \sum_{n\in \IZ} e^{\I \pi \tau (n+\half)^2} \\
\vartheta_3 & = \sum_{n\in \IZ} e^{\I \pi \tau n^2} \\
\vartheta_4 & = \sum_{n\in \IZ} e^{\I \pi \tau n^2}(-1)^n \\
\end{split}
\ee

\section{Lifting Weyl Groups Of Compact Simple Lie Groups}\label{sec:WeylLift}

The lifting of Weyl groups to subgroups of the normalizer $N(T)$ or a maximal torus is well studied
in the mathematical literature and goes back to work of Tits \cite{tits}.
 To state the general problem more formally, let $G$ be a compact Lie
group of rank $r$  and choose a maximal torus $T$ in $G$.  Let $N(T)$ be the normalizer of $T$ in $G$.
As explained in section \ref{sec:Introduction} the Weyl group is defined as $N(T)/T$ and hence
fits in a  short exact sequence
\be \label{shortexact}
 1 \rightarrow T \rightarrow N(T) \xrightarrow{\pi} W \rightarrow 1
\ee
 We say this short exact sequence of groups splits if there is a group homomorphism $W \rightarrow N(T)$ such
that $W \rightarrow N(T) \rightarrow W$ is the identity map on $W$.  When the sequence splits we can use this homomorphism to
define a  subgroup of $N(T)$ isomorphic to $W$ such that the conjugation action of this subgroup on $T$ induces the
Weyl group action on $T$. In general, the sequence \eqref{shortexact} does not split, although there are examples of
groups for which it does. In general, we say that a subgroup $\widetilde W\subset N(T)$ is a \emph{lifting of $W$} if
there is a surjective homomorphism $\pi: \widetilde W \to W$ such that for all $\tilde g \in \widetilde{W}$ and all
$t\in T$, $\tilde g t \tilde g^{-1} = \pi(\tilde g) \cdot t$. There are infinitely many liftings of $W$, but there is a canonical lifting, known
as the Tits lift. If $G$ is the compact simply connected group with Lie algebra $\fg$ then for the Tits lift the Weyl reflections of simple roots lift to order $4$ elements of $G$. In particular,  the Tits lift is never isomorphic to
$W(\fg)$.

It is worth explaining the situation with respect to $SU(2)$ in more detail since much of it carries over to more general $G$.
In $SU(2)$ we can choose the maximal torus $T \simeq S^1$ to consist of the diagonal matrices
\be
\begin{pmatrix} e^{i \alpha} & 0 \\ 0 & e^{-i \alpha} \end{pmatrix}, \qquad \alpha \in \IR
\ee
The normalizer of $T$ then has two connected components. The first component contains the identity and consists of $T$ itself. The second
component consists of the matrices
\be
\begin{pmatrix} 0 & -1 \\ 1 & 0 \end{pmatrix} \cdot T = \begin{pmatrix} 0 & - e^{-i \alpha} \\ e^{i \alpha} & 0 \end{pmatrix}
\ee
Note that, for all $\alpha$ these elements square to $-1$ and are hence of order four.
Thus this makes it clear that there are two elements in $N(T)/T$ and that $N(T)/T$ is isomorphic to $\IZ/2 \IZ$. It is also clear
that there is no homomorphism from the Weyl group $\IZ/2\IZ$ to $N(T)$ because the first component of $N(T)$ has no elements
that act as a Weyl reflection on $T$ and the second component of $N(T)$ has such elements and all such elements have order four.

In order to discuss the general case of a simple Lie algebra $\fg$ with simply connected Lie group $G$ and maximal torus $T$
we   introduce a set of Chevalley-Serre generators: $e_i, f_i, h_i$, $i=1, \dots, r$ satisfying:
\be
\begin{split}
[h_i, h_j ] & = 0 \\
[e_i, f_j] &= \delta_{ij} h_i \\
  [ h_i, e_j] & = C_{ji} e_j\\
[h_i, f_j]&  = - C_{ji} f_j\\
{\rm ad}(e_i)^{1-C_{ji}}(e_j) & = 0  \qquad i \not= j \\
{\rm ad}(f_i)^{1-C_{ji}}(f_j) & = 0  \qquad i \not= j \\
C_{ij}  & := \frac{2 (\alpha_i, \alpha_j)}{(\alpha_j, \alpha_j)} = \alpha_i(h_j) \\
\end{split}
\ee
where $C_{ij}$ is the Cartan matrix of $\fg$ and the simple coroots $h_i$ define a basis of the
Cartan subalgebra $\ft$. For each $i=1,\dots, r$  there is an embedding of $\fs\fl(2)\rightarrow \fg$ defined by $e_i, f_i, h_i$,
$e\to e_i$ etc. where
\be
[e,f] = h \qquad [h,e] = 2 e \qquad [h,f] = -2 f
\ee

For each simple root $\alpha_i$ we have an order $2$ element of $T$ given by
\be
m_i = {\rm exp}(i \pi h_i)
\ee
Tits showed that there is a canonical abelian extension  $\hat W$ of $W$ by an abelian group $\IZ_2^r$ which does embed
in $G$ \cite{tits}. His work has been extended in a number of directions. Our description below is based on \cite{tits,adamhe,CWW,dw,hammer}.
Recall that the action of a Weyl reflection in a root $\alpha$  on an element $h \in t$ of the Cartan subalgebra is
\be
\sigma_\alpha(h) = h - \langle \alpha, h \rangle h_\alpha
\ee
where $h_\alpha$ is the coroot canonically assigned to $\alpha$.
Denoting reflections in the simple roots, $\sigma_{\alpha_i}$,  by $s_i$ we have
\be
s_i(h_j) = h_j - C_{ji} h_i
\ee
where $C_{ji}$ is the Cartan matrix of $G$.

The Weyl group $W$ is generated by the reflections $s_i$, $i=1, \cdots r$.  These obey the relations
\begin{align} \label{weylrel}
s_i^2 &= 1 \\
(s_i s_j)^{m_{i,j}} &=1, \qquad i \ne j \nonumber
\end{align}
where $m_{i,j}$ is the $i,j$ element of the Coxeter matrix. Note that for simple laced $G$ which is our main case of interest $m_{i,j}=2$ if $i \ne j$ and the roots $\alpha_i, \alpha_j$ are orthogonal and $m_{i,j}=3$ if $i \ne j$ and the roots $\alpha_i, \alpha_j$ make
an angle of $2 \pi/3$.

Following \cite{tits} the latter relation can be replaced by
\be
s_i s_j s_i s_j \cdots  s_i s_j = s_j s_i s_j s_i \cdots s_j s_i
\ee
where on the l.h.s. there are $m_{i,j}$ terms $s_i s_j$ and on the r.h.s. there are $m_{i,j}$ terms $s_j s_i$.
This relation follows from  the second relation in (\ref{weylrel}) by successively multiplying the l.h.s. by $s_i^{-1}$,
$s_j^{-1} s_i^{-1}$ and so on and using $s_i^{-1}=s_i$.

Tits shows that the extension $\hat W$ has generators $a_i$, one for each simple reflection which act on
the $h_i$ as
\be
a_i h_j a_i^{-1} = \sigma_{\alpha_i}(h_j)
\ee
and obey the relations
\begin{align}
a_i^2 &= m_i \\
a_i a_j a_i a_j \cdots a_i a_j &= a_j a_i a_j a_i \cdots a_j a_i
\end{align}
where on the l.h.s. there are $m_{i,j}$ terms $a_i a_j$ and on the r.h.s. there are $m_{i,j}$ terms $a_j a_i$.

The $m_i$ generate an abelian $2$-group $T_2$ which is a subgroup of $T$ and the map from $a_i$ to $m_i$ induces
an exact sequence
\be
1 \rightarrow T_2 \rightarrow \hat W \rightarrow W \rightarrow 1
\ee
When $G$ is the simple and simply connected Lie group associated with $\fg$ we can identify $T_2\cong \IZ^r$,
where $r$ is the rank of $G$ with the subgroup of $T$ of points of order two.

In general it appears to be a complicated problem to figure out which conjugacy classes of Weyl group elements have orders
which double when lifted to $\hat W$, but several examples which are relevant to Narian compactifications are discussed in
\cite{adamhe}. We will content ourselves here with a general discussion for $SU(N)$.

\subsection{Example: $G = SU(N)$}\label{app:SUN-Tits}

We consider $SU(N)$ matrices acting on the defining $N$-dimensional representation.
We choose the
standard system of simple roots
and denote the highest weight of the fundamental representation by
$\lambda^1$. Then, up to the action of a diagonal matrix, a weight basis with weights
\be
\lambda^1 , \lambda^1 - \alpha_1 , \lambda^1 - \alpha_1 - \alpha_2, \cdots, \lambda^1 - (\alpha_1 + \cdots + \alpha_{N-1})
\ee
corresponds to the standard Euclidean basis $e_1, \dots, e_N$ of $\IC^N$. Labeling the weight vectors by
$1,2,\dots, N$ the Weyl reflection
$\bar g_i = \sigma_{\alpha_i}$ acts on these weights as the permutation $(i,i+1)$. Therefore, any
lift to $SU(N)$ must have the form:
\be\label{eq:Gen-hatgi}
\hat g_i = \sum_{k\not=i,i+1} z_k^{(i)} e_{k,k} +  ( x_i e_{i,i+1} + y_i e_{i+1,i} ) \qquad i =1, \dots, N-1
\ee
where $x_i, y_i, z^{(i)}_k$ are phases and the $SU(N)$ condition implies
\be\label{eq:constraint}
z_i x_i y_i = -1 \qquad  z_i:= \prod_{k\not= i,i+1} z_k^{(i)} \quad .
\ee
We claim that \underline{any} choice of $x_i,y_i, z_k^{(i)} $ has the correct conjugation properties to
project to an element of the Weyl group:
\be
\hat g_i h_j \hat g_i^{-1} = \begin{cases} h_j & i \not= j, j\pm 1 \\
- h_i & i=j \\
h_{j} + h_{j+1} & j = i-1 \\
h_{j-1} + h_j & j = i + 1\\
\end{cases}
\ee
as one easily checks by direct computation with \eqref{eq:Gen-hatgi} and
$h_{i} = - \frac{\I}{2}( e_{i,i} - e_{i+1,i+1})$.

Note that
\be
\hat g_i^2 = \sum_{k\not=i,i+1} ( z_k^{(i)})^2 e_{k,k} +   ( x_i y_i)( e_{i,i} + e_{i+1,i+1} )
\ee
\be
(\hat g_i \hat g_{i+1})^3 = \sum_{k \not= i,i+1,i+2} (z_k^{(i)} z_k^{(i+1)})^3 e_{k,k} + (x_i y_i z_{i+2}^{(i)})(x_{i+1} y_{i+1} z_i^{(i+1)})
(e_{i,i} + e_{i+1,i+1} + e_{i+2,i+2})
\ee

\textbf{Definition} Let $\widetilde W(x , y , z) \subset N(T)$ be the subgroup of $N(T)$
generated by the elements $\hat g_i$, where the $x_i, y_i, z_k^{(i)}$ are arbitrary phases subject
only to the constraints \eqref{eq:constraint}.

\textbf{Remarks}:

\begin{enumerate}

\item The subgroups $\widetilde W(x,y,z)$ map surjectively to
the Weyl group under the conjugation action.

\item They are finite subgroups iff  $z_k^{(i)}$ and $x_i y_i$ are all roots of unity.

\item
All such subgroups are related by right-multiplication of the generators by suitable elements of $T$.
That is, for any two such groups determined by $(x,y,z)$ and $(x',y',z')$ there are elements
$t_i\in T$ with $\hat g_i' = \hat g_i t_i$.

\item The Tits lift is \
\be \label{tlift}
g_i= \exp[ \frac{\pi}{2}(e_i - f_i) ]
\ee
where $e_i, f_i$ are Serre generators and is given by taking $z^{(i)}_k=1$, $x_i=1$, and $y_i = -1$. According to
\cite{kac} the expression \eqref{tlift} is true in much greater generality than discussed here.

\end{enumerate}

Now let us ask if we can have a subgroup $W(x,y,z)$ isomorphic to $W(\fs\fu(N))$.
Since we want $\hat g_i^2 = 1$ we must choose $z_k^{(i)}\in \{ \pm 1 \}$ as well as
$x_i y_i = 1$. Then the   constraints
\eqref{eq:constraint} show that $z_i=-1$. Therefore we cannot take all $z_k^{(i)}=1$.

Next we need to check the braid relations:
\be
\hat g_i \hat g_{i+1} \hat g_{i} = \hat g_{i+1} \hat g_{i} \hat g_{i+1}
\ee
For order two elements $(\hat g_i \hat g_{i+1})^3$  simplifies to
\be\label{eq:BraidRels}
(\hat g_i \hat g_{i+1})^3 = \sum_{k \not= i,i+1,i+2} (z_k^{(i)} z_k^{(i+1)}) e_{k,k} + ( z_{i+2}^{(i)} z_i^{(i+1)})
(e_{i,i} + e_{i+1,i+1} + e_{i+2,i+2})
\ee
so for a group isomorphic to the Weyl group we need this to be $=1$, putting some further constraint on the $z_k^{(i)}$.

Finally, for $N>3$ we also must check the relations
\be
\hat g_i \hat g_j = \hat g_j \hat g_i \qquad  \qquad \vert i-j \vert > 1
\ee
This is very constraining and shows that $z^{(j)}_i = z^{(j)}_{i+1} $ for  $\vert i - j \vert > 1$.
Therefore
\be
z^{(j)}_1 = \cdots = z^{(j)}_{j-1} = z^{(j)}_-  \qquad \qquad z^{(j)}_{j+2} = \cdots = z^{(j)}_{N} = z^{(j)}_+
\ee
Now combining these constraints with the constraints \eqref{eq:BraidRels} from the braid relations shows that
in fact all
\be
z_k^{(i)} = z
\ee
must have a common value. Since the $z_i=-1$ this common value must be $z=-1$. But this is only
compatible with the second equation in \eqref{eq:constraint} when $N$ is odd.

We conclude that for $N$ odd we can take all $z^{(j)}_k=-1$ for $k \not= j,j+1$ and $x_j = y_j=1$. This gives an explicit
subgroup $W(x,y,z)$ satisfying all the relations. For $N$ even there is no subgroup of $N(T)$ isomorphic
to the Weyl group and the sequence does not split.

\end{document}